\documentclass[11pt,reqno]{article}
\usepackage{amsmath}
\usepackage{amsfonts}
\usepackage{amssymb}
\usepackage{hyperref}
\usepackage{latexsym}
\usepackage[dvips]{graphicx}
\usepackage{epsf}
\usepackage{color}

\allowdisplaybreaks \numberwithin{equation}{section}

\newcommand{\be}{\begin{equation}}
\newcommand{\ee}{\end{equation}}
\newcommand{\bea}{\begin{eqnarray}}
\newcommand{\eea}{\end{eqnarray}}

\newcommand{\f}{\frac}
\newcommand{\p}{\partial}
\newcommand{\na}{\nabla}
\newcommand{\Tr}{{\rm Tr}}

\let\a=\alpha    \let\g=\gamma  \let\d=\delta
\let\z=\zeta        \let\l=\lambda
\let\m=\mu    \let\n=\nu          \let\r=\rho \let\om=\omega
\let\s=\sigma \let\t=\tau      
    
\let\G=\Gamma     
         
  \let\eps=\epsilon

\newcommand{\Gb}{\bar{\Gamma}}

\newcommand{\Cb}{\bar{C}}

\newcommand{\ft}{\tilde{f}}

\newcommand{\Rt}{\tilde{R}}
\newcommand{\Vt}{\tilde{V}}

\newcommand{\phit}{\tilde{\phi}}

\newcommand{\cD}{\mathcal{D}}
\newcommand{\cF}{\mathcal{F}}

\newcommand{\cN}{\mathcal{N}}

\newcommand{\cR}{\mathcal{R}}

\newcommand{\cT}{\mathcal{T}}

\begin{document}

\thispagestyle{empty}
\begin{flushright} \small
AEI-2013-248
\end{flushright}
\bigskip

\begin{center}
  {\LARGE\bfseries   Brans-Dicke theory\\ [1ex]  in the local potential approximation}\\
[10mm]
{\large Dario Benedetti${}^a$ and Filippo Guarnieri${}^{a,b}$}
\\[3mm]
{\small\slshape
${}^a$ Max Planck Institute for Gravitational Physics (Albert Einstein Institute), \\
Am M\"{u}hlenberg 1, D-14476 Golm, Germany \\ 
${}^b$ Dipartimento di Fisica, Universit$\grave{a}$ degli Studi di Roma Tre and\\ 
INFN sezione di Roma Tre, Via della Vasca Navale 84, I-00146 Rome, Italy\\
\vspace{.3cm}
 {\upshape\ttfamily dario.benedetti@aei.mpg.de, filippo.guarnieri@aei.mpg.de} 
} 

\end{center}
\vspace{5mm}

\hrule\bigskip

\centerline{\bfseries Abstract} \medskip

We study the Brans-Dicke theory with arbitrary potential within a functional renormalization group framework. Motivated by the asymptotic safety scenario of quantum gravity and by the well-known relation between $f(R)$ gravity and Brans-Dicke theory at the classical level, we concentrate our analysis on the fixed-point equation for the potential in four dimensions and with Brans-Dicke parameter $\omega = 0$. 
For two different choices of gauge, we study the resulting equations by examining both local and global properties of the solutions, by means of analytical and numerical methods.
As a result of our analysis we do not find any nontrivial fixed point in one gauge, but we find a continuum of fixed points in the other one. We interpret such inconsistency as a result of the restriction to $\omega=0$, and thus we suggest that it indicates a failure of the equivalence between $f(R)$ gravity and Brans-Dicke theory at the quantum level.

\bigskip
\hrule\bigskip
\newpage
\tableofcontents

\section{Introduction}
\label{Sec:intro}

Many models of modified gravity have been proposed and studied over time in an attempt to address the problems encountered in quantum gravity and cosmology  \cite{Clifton:2011jh}.
New models are most commonly postulated as the starting point of a new research direction, however, one instance in which the logic is partially reversed, and a new model of gravity is hoped to emerge as the final result, is the asymptotic safety scenario \cite{Weinberg:1980gg,Niedermaier:2006wt,Percacci:2007sz,Litim:2011cp,Reuter:2012id}. The general idea behind such scenario is that the theory of quantum gravity should be sought within a large class of theories (e.g. all possible theories described by an action functional of a single metric field), out of which one single theory (or few isolated ones) should emerge with the peculiar characteristic of being a fixed-point of the renormalization group (RG) flow. IR-unstable trajectories emanating from such fixed point(s) would then define nonperturbatively renormalized theories of gravity. The use of functional renormalization group equations (FRGEs) has provided considerable evidence in support of the existence of a nontrivial fixed point theory for gravity, in a large number of formulations and approximations (see list of references in \cite{Litim:2011cp,Reuter:2012id}).

Even within a given class of models, specified by a choice of variables and symmetries, it is obviously impossible to explore the entire space of theories and one has to resort to approximations. A common approximation in the asymptotic safety literature consists in truncating the theory space to a finite-dimensional subspace by making a guess of what might be the most important terms to keep track of in  the effective action. The guess is then supposed to be repeatedly refined until little improvement of the results is gained by new refinements. In practice, even this is quite hard, and only recently such program has been implemented to a high order of refinement for truncations that only retain polynomials of the Ricci scalar \cite{Codello:2007bd,Machado:2007ea,Codello:2008vh,Bonanno:2010bt,Falls:2013bv}.
At the same time, the functional nature of the renormalization group methods being used has just started being exploited further, opening up the possibility of exploring infinite-dimensional subspaces of the theory space.
The main idea is to study the RG flow of gravity in a spirit similar to the derivative expansion of scalar field theory \cite{Morris:1998da,Bagnuls:2000ae,Berges:2000ew,Delamotte-review}. There, the leading order approximation is known as local potential approximation (LPA), and it consists in projecting the flow equation on a constant scalar field, thus allowing to study the flow of a generic effective potential $V_k(\phi)=\G_k[\phi={\rm const.}] /  \int d^d x $. The next to leading order includes a term with two derivatives and any field dependence, and so on at higher orders. At each order of the derivative expansion one is lead to study partial differential equations for functions of the field $\phi$ and of the running scale $k$.
In gravity there is much more structure, and there are probably many options in organizing an expansion of this sort. A very natural option is to organize the action as if it was an expansion around a maximally symmetric background. For the latter the only non-zero component of the Riemann tensor is the Ricci scalar $R$, which is constant: we have $\na_\m R = S_{\m\n} = C_{\m\n\r\s}=0$, where $S_{\m\n}$ is the traceless Ricci tensor, and $C_{\m\n\r\s}$ is the Weyl tensor. The analogue of the derivative expansion can then be an expansion in $S_{\m\n}$, $C_{\m\n\r\s}$ and their derivatives (by the Bianchi identity $\na_\m R =\f{2d}{d-2} \na^\n S_{\m\n}$), with arbitrary dependence on $R$ at each order. In the leading order of such an expansion we are left with an $f(R)$ theory, whose study in such spirit was begun in \cite{Benedetti:2012dx,Demmel:2012ub,Dietz:2012ic,Benedetti:2013jk,Dietz:2013sba}.

As compared to the LPA for scalar field theory, in the $f(R)$ approximation for gravity we face a number of additional technical complications, in particular a larger number of contributions to the FRGE, with a more complicated dependence on the unknown function, and the challenge of evaluating functional traces on a curved background. The latter in particular introduces some subtleties related to the presence of zero modes in compact backgrounds and to the staircase nature of the results obtained for the traces when using cutoffs with step functions \cite{Benedetti:2012dx}.
Also for these reasons progress has been slow in this direction, and it is desirable to find alternative ways to study the same problem.

One possible simplification, which we will explore in this paper, is suggested by a well known classical property of the $f(R)$ theory \cite{Sotiriou:2008rp,DeFelice:2010aj}.
The classical action for $f(R)$ gravity,\footnote{As in most asymptotic safety research (for a rare exception see \cite{Manrique:2011jc}), we will be working in Euclidean signature.}
\be \label{fR-action}
S[g] =  \int d^d x \sqrt{g}\,  f(R) \, ,
\ee
can be traded for an equivalent action, describing a scalar-tensor theory,
\be \label{phi-action}
S[g,\phi] =  \int d^d x \sqrt{g}  \left \{ V(\phi)-\phi \, R \right \} \, .
\ee
The relation between the two Lagrangians is given by a Legendre transform,
\be
V(\phi)= R(\phi)\, \phi + f(R(\phi))\, ,
\ee
where $R(\phi)$ is found by solving the equation $\phi=-f'(R)$ for $R$, and as usual, regularity of the transform is guaranteed if $f''(R)\neq 0$.
From a RG perspective the advantage of such formulation is that we can study the running of the potential by projecting the FRGE on  a flat background, thus sidestepping all the complications of curved backgrounds.
In fact, we will construct flow equations for a more general version of \eqref{phi-action}, that is, a generic Brans-Dicke theory with a potential (see \eqref{BD-action}).\footnote{Note that in the context of asymptotic safety, Brans-Dicke theory with $\om=0$ was considered in \cite{Reuter:2003ca} as a RG improvement of the Einstein-Hilbert truncation, in which the running gravitational and cosmological constants were promoted to fields as a result of an identification of scales with spacetime points. Clearly our work differs substantially from  \cite{Reuter:2003ca}, as we study the RG equations directly for the Brans-Dicke theory. In a sense our work relates to  \cite{Reuter:2003ca} like the general $f(R)$ studies \cite{Benedetti:2012dx,Demmel:2012ub,Dietz:2012ic,Benedetti:2013jk,Dietz:2013sba} relate to the $f(R)$ actions obtained by improvement of the Einstein-Hilbert truncation \cite{Bonanno:2012jy,Hindmarsh:2012rc,Domazet:2012tw}.}
Projection on a flat background will allow us to study such theory without truncating the potential to a polynomial form, thus performing an analysis similar to that of pure scalar theory \cite{Hasenfratz:1985dm,Zumbach:1994vg,Morris:1994ie,Morris:1994ki,Bervillier:2007rc,Codello:2012sc}.

Of course, at a quantum level the two theories might be inequivalent. 
They are both perturbatively nonrenormalizable, and standard perturbative reasonings could only apply at an effective field theory level. When looking for a UV completion in the form of a nontrivial fixed point, we study the RG equations in two different theory spaces, and in the full fixed point theory the scalar field might couple to other geometric invariants, or acquire its own dynamical term. Moreover, since the functions $f(R)$ or $V(\phi)$ are not chosen a priori, but have to be found such that they correspond to an RG fixed point, it could happen that the regularity of the transform fails at one or several field values (or even in the full range of definition if for example $f(R)$ is linear at the fixed point).
As a consequence, if fixed points were to be found in both formulations, they might describe different physics. It might also happen that one formulation admits an asymptotic safety scenario and the other does not.\footnote{In addition, we should also notice that often in the cosmology literature other ``frames'' are considered, in which a new metric field is defined via a conformal map, often together with a redefinition of the scalar field as well. Again, at the classical level these are all equivalent theories (although there has been some confusion on the issue in the past \cite{Flanagan:2004bz}), but at the nonperturbative quantum level this is not guaranteed (perturbatively there is on-shell equivalence, as shown for example in two dimensions \cite{Elizalde:1993qq}, but at the nonperturbative level this has not been shown, although it might be possible following the developments of \cite{Percacci:2011uf,Codello:2012sn}).
We will not study here other frames, having always in mind the original pure metric theory, whose metric we assume to define the coupling to ordinary matter.}
However, we also do not know a priori whether the (nonperturbatively renormalized) quantum theories are equivalent or not, and only a direct comparison (which we can at least do at the level of truncations) could allow us to settle the question. 
In any case, given that in asymptotic safety we are in principle allowing for extra degrees of freedom, there seems to be no reason to consider only pure metric theories of gravity, and the study of scalar-tensor theories is of interest in its own. Brans-Dicke theory is  one of the oldest modifications of general relativity \cite{Brans:1961sx}, and together with its variations and generalizations it finds plenty of applications in cosmology \cite{Clifton:2011jh}, and in quantum gravity (e.g. \cite{Nojiri:2000ja,Grumiller:2002nm,Zhang:2011vg}).
Other versions of scalar-tensor theories have already been studied in the context of asymptotic safety \cite{Narain:2009fy,Narain:2009gb,Henz:2013oxa}, but to the best of our knowledge,  no study of this sort has been done before for the Brans-Dicke theory in the formulation we consider here (sometimes referred to as Helmholtz-Jordan frame).

We will introduce more precisely our ansatz and setup in Sec.~\ref{Sec:setup}, together with the two choices of gauge fixing that we are going to employ.
In Sec.~\ref{Sec:frge} we will derive the FRGEs for both gauges for general dimension and Brans-Dicke parameter, while  in Sec.~\ref{Sec:anal} we will proceed to analyze their local properties in $d=4$ and $\om=0$.
Finally, we will present the results of numerical integrations in Sec.~\ref{Sec:fps}, concluding in Sec.~\ref{Sec:concl} with a discussion of our findings and of future prospects.
In App.~\ref{App:2d} we analytically solve the special case $d=2$, which helps illustrating some of the points discussed in the conclusions.

\section{The general setup: ansatz, variations and gauge fixing}
\label{Sec:setup}

The action \eqref{phi-action} is a particular case ($\om=0$) of the more general Brans-Dicke theory with a potential,
\be \label{BD-action}
\Gb_k[g,\phi] =  \int d^d x \sqrt{g} \left \{ V_k(\phi) - \phi\, R+ \f{\om}{\phi}\, \p_\m\phi\, \p^\m\phi \right \} \, ,
\ee
which in turn is a special case of dilaton gravity.
The $f(R)$ theory in the Palatini formalism is related to the same theory but with $\om=-3/2$ \cite{Sotiriou:2006hs}. 
We have introduced a subscript $k$ which stands for the running scale at which the effective average action $\Gb_k$ is defined \cite{Berges:2000ew}.
To a large extent we will keep $\om$ general, only to concentrate on the specific case $\om=0$ for our numerical analysis (studying the running of $\om$ would require using a non-constant background, or looking at the 2-point function, which we leave for future work).
Note that \eqref{BD-action} differs from other scalar-tensor theories studied in the asymptotic safety literature \cite{Narain:2009fy,Narain:2009gb,Henz:2013oxa} in two important aspects: it is not invariant under $\phi\to-\phi$ (and of course $\phi$ is not restricted to be positive), and the kinetic term (when present, that is, when $\om\neq 0$) contains an inverse of the field.

The point of view we wish to adopt in this paper is that the action \eqref{BD-action} is a LPA approximation for the effective action, and that only to next order we would promote $\om$ and the function coupled to $R$ to general functions of $\phi$.
As we explained in the introduction, we will project the flow equation for \eqref{BD-action} on a flat background and study only the running of the potential. However, for future reference, we will present in this section the results of variations and gauge fixing for a general maximally symmetric background metric and constant background scalar field.

Introducing the background splitting
\be
g_{\m\n}\to g_{\m\n}+h_{\m\n}\, , \;\;\;\;\; \phi\to\phi+\varphi\, ,
\ee
we make the usual approximation for the effective average action \cite{Reuter:1996cp}
\be
\G_k[h,\varphi;g,\phi] = \Gb_k[g+h,\phi+\varphi] + S_{\rm gf}[h,\varphi;g,\phi] +S_{\rm gh}[\Cb,C,h,\varphi;g,\phi]\, ,
\ee
and neglect the running of the gauge-fixing and ghost actions, $S_{\rm gf}$ and $S_{\rm gh}$.

For the FRGE we will need the second variation of the effective average action, therefore we expand 
\be
\Gb[g+\eps\, h,\phi+\eps\, \varphi]=\Gb[g,\phi]+\eps\, \d^{(1)}\Gb[h,\varphi;g,\phi]+\eps^2 \d^{(2)}\Gb[h,\varphi;g,\phi]+O(\eps^3)\, ,
\ee
and find (omitting from now on the field dependencies of the action functionals)

\be
\begin{split}
\d^{(2)}\Gb_k =  &  \int   d^d x \,\sqrt{g}\, \Big\{\varphi\, \left(-\f{\om}{\phi}\na^2 + \f12 V_k''(\phi)\right)\, \varphi \\
  & +\varphi\, \left(\na^2 h-\na^\m\na^\n h_{\m\n}+\f{2-d}{2d}R\, h+\f12 V_k'(\phi)\, h\right) + V_k(\phi)\left(\f18 h^2-\f14 h^{\m\n}h_{\m\n}\right)\\
  & +  \f18 \phi\, h_{\m\n} \Big( -(g^{\m\r}g^{\n\s}+g^{\m\s}g^{\n\r}-2 g^{\m\n}g^{\r\s})\na^2 + 4\, g^{\r\m}\na^{\n}\na^\s - 4\, g^{\r\s}\na^{\m}\na^\n \Big)h_{\r\s} \\
   & +  \phi\, R\, \Big( \f{d\,(d-3)+4}{4\,d\,(d-1)} h^{\m\n}  h_{\m\n} - \f{d\,(d-5)+8}{8\,d\,(d-1)} h^2\Big)\Big\} \, .
\end{split}
\ee
We can exploit the gauge-fixing freedom to simplify the Hessian operator, adding to the original action the gauge-fixing term
\be
S_{\rm gf} = \f{1}{2\, \a}  \int   d^d x \, \sqrt{g}\,  \cF_\m G^{\m\n} \cF_\n \, ,
\ee
for some choice of gauge-fixing constraint $\cF_\m$ and of non-degenerate operator $G^{\m\n}$.
Physical results should be independent of the gauge choice, however, it is well known that the off-shell effective action is not gauge independent, and furthermore, the approximations we employ in the FRGE lead to additional gauge dependences. It is then important to test our analysis against different choices of gauge.
We present in the following the two types of gauge which we will use in the forthcoming sections.

\subsection{Feynman gauge}
First we consider a Feynman gauge ($\a=1$) with
\be
\cF^{(F)}_\m =  \na^\n \left(h_{\m\n} -\f12 g_{\m\n} h\right) - \f{1}{\phi}\, \na_\m \varphi \, ,
\ee
and
\be
G^{(F)}{}^{\m\n} = \phi\, g^{\m\n}\, .
\ee

The total quadratic action becomes

\be
\begin{split}
\d^{(2)}\Gb_k +S^{(F)}_{\rm gf} =  &  \int   d^d x \, \sqrt{g}\, \Big\{\f12 \varphi \Big(-\f{1+2\,\om}{\phi}\na^2+V_k''(\phi) \Big) \varphi \\
 &+\f12 \varphi \Big(\na^2+\f{2-d}{d}R+V_k'(\phi)\Big) h \\
 & -  \f18 h_{\m\n} \Big( (g^{\m\r}g^{\n\s}+g^{\m\s}g^{\n\r}- g^{\m\n}g^{\r\s})(\phi\, \na^2 +V_k(\phi))  \Big)h_{\r\s}\\
 & +  \phi\, R\, \Big( \f{d\,(d-3)+4}{4\,d\,(d-1)} h^{\m\n}  h_{\m\n} - \f{d\,(d-5)+8}{8\,d\,(d-1)} h^2\Big) \Big\} \, .
\end{split}
\ee
Decomposing $h_{\m\n}=\hat{h}_{\m\n}+\f1d g_{\m\n} h$, with $g^{\m\n}\hat{h}_{\m\n}=0$, we finally obtain
\be \label{HessFeyn}
\begin{split}
\d^{(2)}\Gb_k +S^{(F)}_{\rm gf} =  &  \int   d^d x\,\sqrt{g}\,  \Big\{\f12 \varphi \Big(-\f{1+2\om}{\phi}\na^2+V_k''(\phi) \Big) \varphi \\ 
 & +\f12 \varphi \Big(\na^2+\f{2-d}{d}R+V_k'(\phi)\Big) h \\ 
  & -  \f14 \hat{h}^{\m\n}  \Big(\phi\, \na^2 -  \f{d\,(d-3)+4}{d\,(d-1)}\, \phi\, R +V_k(\phi)\Big)  \hat{h}_{\m\n} \\ 
  & + \f{d-2}{8\,d} h\, \Big(\phi\, \na^2 - \f{d-4}{d}\, \phi\, R +V_k(\phi) \Big) h\Big\} \, . 
\end{split}
\ee
We note that via the gauge-fixing procedure we have introduced a kinetic term for the auxiliary field $\varphi$ even in the case $\om=0$. The kinetic term disappears for $\om=-1/2$, which is a special value for the Brans-Dicke theory in this gauge.

For the gauge sector we employ a standard Fadeev-Popov determinant which we rewrite in terms of a quadratic integral over complex Grassmann fields $C^\mu$ and $\bar{C}^\mu$. For constant background scalar field, the ghost action reads
\be
\label{ghostsector}
S_{gh}[C,\bar{C}] = \int\, d^d x\, \sqrt{g}\, \left \{ \bar{C}^\mu\, \left(\nabla^2 + \f{R}{d}\right)\, C_\mu \right \} \,.
\ee

\subsection{Landau gauge}
As an alternative choice of gauge, we consider a Landau gauge ($\a=0$) with
\be
\cF^{(L)}_\m =  \na^\n\, \left(h_{\m\n} -\f1d\, g_{\m\n}\, h\right)  \, ,
\ee
and
\be
G^{(L)}{}^{\m\n} =  g^{\m\n}\, .
\ee
The interesting aspect of such gauge is that it does not modify the kinetic term of $\varphi$, in particular it does not introduce one for $\om=0$.

In this case, in order to simplify the non-minimal operators that appear in the second variation, we use the transverse-traceless decomposition of the metric fluctuations, given by
\be \label{TT-dec}
h_{\m\n} = h_{\m\n}^{T} + \na_\m \xi_\n + \na_\n \xi_\m + \na_\m \na_\n \s + \frac{1}{d} g_{\m\n} (h-\na^2 \s)\, ,
\ee
with the component fields satisfying
\be
g^{\m\n} \, h_{\m\n}^{T} = 0 \, , \quad \na^\m h_{\m\n}^{T} = 0
\, , \quad \na^\m \xi_\m = 0 \, , \quad h=g_{\m\n} h^{\m \n} \, .
\ee

In the $\a\to 0$ limit, the $\xi_\m$ and $\s$ field components decouple completely from the rest of the Hessian, and their contribution to the FRGE cancels exactly with the ghost contribution, when properly implemented \cite{Benedetti:2011ct}.
We thus write the second variation of the action directly omitting the contribution of the longitudinal components:
\be \label{HessLand}
\begin{split}
\d^{(2)}\Gb_k +S^{(L)}_{\rm gf} =  &  \int   d^d x\,\sqrt{g}\,  \Big\{\f12 \varphi \Big(-\f{2\, \om}{\phi}\na^2+V_k''(\phi) \Big) \varphi \\
 & + \varphi \Big(\f{d-1}{d} \na^2+\f{2-d}{2\,d}R+\f12 V_k'(\phi)\Big) h \\
  & -  \f14 h^{T}{}^{\m\n}  \Big(\phi\, \na^2 - \f{d\,(d-3)+4}{d\,(d-1)}\, \phi\, R +V_k(\phi)\Big)  h_{\m\n}^{T} \\ 
  & + \f{d-2}{8\,d} h\, \Big(2\f{d-1}{d}\, \phi\, \na^2 -\f{d-4}{d}\, \phi\, R +V_k(\phi) \Big) h\Big\} \, .
\end{split}
\ee

Because of the change of variables \eqref{TT-dec}, in this case there is also a Jacobian to keep track of, which we do by introducing auxiliary fields. 
The Jacobian for the gravitational sector leads to the auxiliary action \cite{Benedetti:2011ct}
\be \label{aux-gr}
\begin{split}
S_{\rm aux-gr} =  \int d^d x \sqrt{g}\, \Big\{ & 2\, \bar{\chi}^{T\, \m} \Big( \na^2 +\f{R}{d}\Big) \chi_\m^T + \Big( \f{d-1}{d}\Big) \bar{\chi} \Big( \na^2 +\f{R}{d-1}\Big) \na^2 \chi    \\
   & + 2\, \z^{T\m} \Big( \na^2 +\f{R}{d}\Big) \z_\m^T +\Big( \f{d-1}{d}\Big) \z \Big( \na^2+ \f{R}{d-1}\Big) \na^2 \z  \Big\}\, ,
\end{split}
\ee
where the $\chi_\m^T$ and $\chi$ are complex Grassmann fields, while $\z_\m^T$ and $\z$ are real bosonic fields.
The Jacobian for the transverse decomposition of the ghost action is given by
\be
S_{\rm aux-gh} = \int d^d x \sqrt{g} \, \psi \na^2 \psi \, ,
\ee
with $\psi$ a real scalar field.

\section{The flow equation}
\label{Sec:frge}
The flow equation can be evaluated by means of the Functional Renormalization Group Equation (FRGE), which takes the generic form \cite{Morris:1998da,Bagnuls:2000ae,Berges:2000ew,Delamotte-review}
\be
\label{frge}
\partial_t\, \Gamma_k[\Phi]  = \frac{1}{2} \text{STr} \left[ \left( \Gamma_k^{(2)} + \mathcal{R}_k \right)^{-1}\, \partial_t\, \mathcal{R}_k \right ] \, ,
\ee
being
\be
\Gamma_k^{(2)}(x,y) = \frac{\d^2 \Gamma_k}{\d\Phi_i(x)\, \d\, \Phi_j(y)} \, ,
\ee
and where $\Phi$ is a superfield collecting all the fields involved in the quantum action, i.e. $\Phi \equiv \{\varphi, h_{\mu\nu}, \cdots\}$. $\mathcal{R}_k$ is a generic cutoff operator, $t \equiv \log(k)$ is the RG running scale and STr identifies a functional supertrace, carrying a factor 2 for complex fields and a factor $-1$ for Grassmann fields.

We will construct the cutoff operator in such a way to implement the substitution rule 
\be \label{cutrule}
-\na^2\;\; \to\;\; P_k \equiv -\na^2 + k^2 r(-\na^2/k^2) \, ,
\ee
being $r(z)$ a dimensionless smearing function. That is, we choose a cutoff of the form $\cR_k=\G_k^{(2)} {|}_{-\na^2\to P_k}-\G_k^{(2)}$.
A convenient choice of  smearing function, leading to a considerable simplification of the functional traces, and which we will therefore use here, is the so-called ``optimized" cutoff \cite{Litim:2001up} which reads
\be
\label{optimized}
r(z) = (1 - z) \Theta(1 - z)\, ,
\ee
where $\Theta(x)$ is a Heaviside step function.

\subsection{Feynman gauge}

The Hessian of the effective action is mostly diagonal in field space, with the only exception of the $\{h, \varphi\}$ sector,
thus the supertrace in \eqref{frge} can be easily decomposed into standard functional traces.
In the Feynman gauge we obtain
\be
\label{blockFeynman}
\partial_t\, \Gamma_k[\Phi] =  \frac{1}{2} \Tr \left[ \left(\mathcal{H}_k \right)^{-1}\, \partial_t\, \mathcal{R}_k \right ]_{h, \varphi}  +
\frac{1}{2} \Tr \left[ \left(\mathcal{H}_k \right)^{-1}\, \partial_t\, \mathcal{R}_k \right ]_{h^T, h^T} -
 \Tr \left[  \left(\mathcal{H}_k \right)^{-1}\, \partial_t\, \mathcal{R}_k \right ]_{\Cb,C}\, ,    
\ee
where $\mathcal{H}_k$ is the modified inverse propagator, namely $\mathcal{H}_k =  \Gamma_k^{(2)} + \mathcal{R}_k$.
The evaluation of the first trace requires to invert the $h$-$\varphi$ matrix, which is trivial since the matrix elements commute. The ghost term takes a factor of minus two with respect to the other terms, because of the complex Grassmannian nature of the ghost fields.

The trace over a generic Riemannian manifold can be evaluated by means of a heat kernel expansion, but since we are interested in projecting the flow equation on a flat background we can evaluate the trace over modes as a simple integral over momenta. 
The derivative of the cutoff operator with respect to the RG time returns
\be
\partial_t\, k^2 r\left(\f{p^2}{k^2}\right) = 2\, k^2\, \Theta\left(1 - \f{p^2}{k^2}\right) +2\, \f{p^2}{k^2}\,  (k^2 - p^2)\, \delta\left(1 - \f{p^2}{k^2}\right)\, ,
\ee
which reduces to the sole Heaviside step function using the property that the distributional product of the delta function with its argument is zero. Because of the step function, the trace reduces to a momentum integral between $0$ and $k$, thus automatically rendering the functional traces UV finite, a well-known feature of the FRGE. Performing the trace we obtain
\be
\frac{1}{2} \Tr \left[ \left(\mathcal{H}_k \right)^{-1}\, \partial_t\, \mathcal{R}_k \right ]_{h, \varphi} =  \frac{2^{1-d} \pi^{-\frac{d}{2}}}{d\, \Gamma\left(\frac{d}{2}\right) } k^{d+2} \frac{N_F}{D_F}
\ee
\be
\frac{1}{2} \Tr \left[ \left(\mathcal{H}_k \right)^{-1}\, \partial_t\, \mathcal{R}_k \right ]_{h^T, h^T} =
 \left(\frac{d\,(d+1)}{2} -1\right)\,\frac{2^{1-d}\, \pi^{-\frac{d}{2}}}{d\, \Gamma\left(\frac{d}{2}\right)}  k^{d+2} \,\frac{\phi}{\left(k^2 \phi - V(\phi)\right)}
\ee
\be
\Tr \left[  \left(\mathcal{H}_k \right)^{-1}\, \partial_t\, \mathcal{R}_k \right ]_{\Cb,C} = \frac{2^{2-d} \pi^{-\frac{d}{2}}}{\Gamma\left(\frac{d}{2}\right)}\,  k^d
\ee
where
\bea
N_F &=&  \phi \left \{4\, k^2 (d\, \omega +d-2\, \omega -1) + (d-2)\, \phi\, V''(\phi) - 2\, d\, V'(\phi) \right \} + \nonumber \\ 
& & (2-d) (2\, \omega +1) V(\phi) \, , \nonumber \\[1em]
D_F &=&(2-d) V(\phi) (k^2\, (2\, \omega+1)+\phi\, V''(\phi))+ \phi\, \big \{k^2 (2\, k^2 (d\, \omega+d-2 \omega -1) +  \nonumber \\
& & (d-2)\, \phi\, V''(\phi)) - 2\, d\, k^2 V'(\phi) + d\, V'(\phi)^2 \big \}  \, \nonumber .
\eea
The trace over the tensor structure gives the factor  $d(d+1)/2 -1$ for the $h^T_{\m\n}$ contribution and a factor $d$ for the ghosts, counting the number of their independent components.
Since we are working on a flat manifold and constant background field both the Ricci scalar and the kinetic operator vanish, so that equation (\ref{blockFeynman}) reduces to an RG flow equation for the dimensionful potential.  
We cast the equation in an autonomous form, i.e. with no explicit dependence on $k$, by introducing the dimensionless quantities
\be \label{dimless}
\phit = \phi \, k^{2-d}\, , \hspace{0.2cm}  \Vt(\phit) = V(k^{d-2}\phit) \, k^{-d}\, ,
\ee
in terms of which we obtain
\be \label{RGFeyn}
\partial_t\, \Vt_k(\phit) = \cT_{\rm tree} + \cT_{\rm quant}^{(F)}\, ,
\ee
where
\be \label{RG-tree}
 \cT_{\rm tree} =  - d\, \Vt(\phit) + (d - 2)\, \phit  \, \Vt'(\phit)\, ,
\ee
is the classical part of the equation, which is linear in the potential, and
\be \label{RGFeyn-quant}
\cT_{\rm quant}^{(F)} =  \frac{2^{1-d} \pi^{-\frac{d}{2}}}{d\, \Gamma\left(\frac{d}{2}\right) } \left \{   -2\, d    + \frac{ \left(\frac{1}{2} \left(d^2+d\right)-1\right)\, \phit}{\left( \phit - \Vt(\phit)\right)} + \frac{\widetilde{N}_F}{\widetilde{D}_F} \right \} \, ,
\ee
with
\bea
\widetilde{N}_F &=& \phit \left \{4\, (d\, \omega + d - 2 \omega -1) + (d-2)\, \phit\, \Vt''(\phit) - 2\, d\, \Vt'(\phit) \right \} + \nonumber \\ 
& & (2-d) (2\, \omega +1) \Vt(\phit) \, , \\[1em]
\widetilde{D}_F &=&(2-d) \Vt(\phit) ((2 \,\omega+1)+\phit \Vt''(\phit)) + \phit \, \big \{(2\, (d\, \omega + d - 2 \omega -1) + \nonumber \\
& & (d-2)\, \phit\, \Vt''(\phit)) - 2\, d\, \Vt'(\phit) + d\, \Vt'(\phit)^2 \big \}  \, ,  \nonumber
\eea
is the quantum part, which contains all the loop contributions, and which is responsible for the nonlinear character of the equation.

\subsection{Landau gauge}
Working in the Landau gauge  the supertrace in \eqref{frge} reads
\be
\label{blockLandau}
\partial_t\, \Gamma_k[\Phi] =  \frac{1}{2} \Tr \left[ \left(\mathcal{H}_k \right)^{-1}\, \partial_t\, \mathcal{R}_k \right ]_{h, \varphi}  +
\frac{1}{2} \Tr \left[ \left(\mathcal{H}_k \right)^{-1}\, \partial_t\, \mathcal{R}_k \right ]_{h^{TT}, h^{TT}} +
\frac{1}{2} \text{STr} \left[  \left(\mathcal{H}_k \right)^{-1}\, \partial_t\, \mathcal{R}_k \right ]_{aux}    ,
\ee
where the contributions of ghosts and longitudinal modes have been omitted, since they exactly cancel each other as explained before.

After performing the integral over momenta we obtain
\be
\frac{1}{2} \Tr \left[ \left(\mathcal{H}_k \right)^{-1}\, \partial_t\, \mathcal{R}_k \right ]_{h, \varphi} =  \frac{2^{2-d} \pi^{-\frac{d}{2}}}{d\, \Gamma\left(\frac{d}{2}\right) } k^{d+2} \frac{N_L}{D_L} \, ,
\ee
\be
\frac{1}{2} \Tr \left[ \left(\mathcal{H}_k \right)^{-1}\, \partial_t\, \mathcal{R}_k \right ]_{h^{TT}, h^{TT}} =
 \left(\frac{d\,(d+1)}{2}-d-1\right)\, \frac{2^{1-d} \pi^{-d/2}}{d\, \Gamma \left(\frac{d}{2}\right)}\, k^{d+2}\, \frac{\phi}{\left(k^2 \phi - V(\phi)\right)} \, ,
\ee
\be
\frac{1}{2} \text{STr} \left[ \left(\mathcal{H}_k \right)^{-1}\, \partial_t\, \mathcal{R}_k \right ]_{aux} =
-\frac{2^{1-d} \pi ^{-d/2}}{\Gamma \left(\frac{d}{2}\right)} \, k^d \, ,
\ee
\bea
N_L &=& (d-1)\, \phi \left \{4\, k^2 (d\, \omega  + d - 2 \, \omega -1) + ( d - 2 )\, \phi\, V''(\phi) - 2\, d\, V'(\phi) \right \} + \nonumber \\ 
& & ( 2 - d ) \,d\, \omega\,  V(\phi) \, , \nonumber \\[1em]
D_L &=& \phi\, \large \{d^2\, V'(\phi)^2 + 2\, (d - 1)\, k^2 \left(2\, k^2 (d\, \omega +d-2\, \omega -1)+(d-2)\, \phi\, V''(\phi) \right) + \nonumber  \\ 
& & -4\, (d-1)\, d\, k^2 V'(\phi) \large \} + ( 2 - d )\, d\, V(\phi) (2\, k^2 \omega +\phi\, V''(\phi)) \, \nonumber.
\eea
The RG flow equation for the dimensionless potential in such a gauge reads then
\be \label{RGLand}
\partial_t\, \Vt_k(\phit) = \cT_{\rm tree} + \cT_{\rm quant}^{(L)}\, ,
\ee
where the classical part $\cT_{\rm tree}$ is the same as in \eqref{RG-tree}, and the quantum part reads
\be
\cT_{\rm quant}^{(L)} = \frac{2^{1-d} \pi^{-\frac{d}{2}}}{d \,\Gamma\left(\frac{d}{2}\right) } \left \{   -d    + \frac{ \left(\frac{1}{2} \left(d^2+d\right)-d-1\right)\, \phit}{\left( \phit - \Vt(\phit)\right)} + 2\, \frac{\widetilde{N}_L}{\widetilde{D}_L} \right \} \, ,
\ee
with
\bea
\widetilde{N}_L &=& (d-1) \,\phit \left \{4 \, (d\, \omega  + d - 2\,  \omega -1) + ( d - 2 )\, \phit \,\Vt''(\phit) - 2\, d\, \Vt'(\phit) \right \} + \nonumber\\ 
& &  ( 2 - d )\, d\, \omega\,  \Vt(\phit) \, , \\[1em]
\widetilde{D}_L &=& \phit\, \large \{d^2\, \Vt'(\phit)^2 + 2\, (d - 1)\, \left(2\, (d\, \omega + d - 2\,  \omega -1) + (d-2)\, \phit \,\Vt''(\phit) \right) + \nonumber \\ 
& & - 4\, (d-1)\, d\, \Vt'(\phit) \large \} + ( 2 - d )\, d\, \Vt(\phit) (2\, \omega + \phit \, \Vt''(\phit)) \, \nonumber .
\eea

\section{Analytical study of the equation}
\label{Sec:anal}

We want to search for fixed point solutions of equation (\ref{RGFeyn}) and (\ref{RGLand}), i.e. search for scale invariant solutions such that $\partial_t\, \Vt_k = 0$, requiring them to be globally analytic \cite{Hasenfratz:1985dm,Felder:1987,Morris:1994ki}. The latter requirement has a well-understood physical and mathematical justification, being a necessary condition for the existence of the average effective action at all values of $k$, and hence of the full effective action in the limit $k\to 0$ (which in $d>2$ requires the existence of the solution for $\phit\to\pm\infty$, see \eqref{dimless}). In addition, the condition of global analyticity is expected to reduce the continuous set of solutions to a discrete subset of acceptable ones.

For $\partial_t\, \Vt_k = 0$, both  (\ref{RGFeyn}) and (\ref{RGLand}) reduce to second order ordinary differential equations, thus we expect 2-parameter families of local solutions, parametrized by the initial value conditions. Extending such local solutions to global ones, we generally have to impose constraints coming from the analyticity requirement and from the symmetries of the problem.
In our case we do not have any constraints originating from symmetries (e.g. we have no $\phit\to -\phit$ symmetry, hence $\Vt'(0)\neq 0$ in general), and we will have to study the equation on the full real line imposing asymptotic boundary conditions at $\phit\sim \pm\infty$. The latter, due to the non-linear nature of the equations, could contain less than two free parameters, implying that global solutions would also necessarily be parametrized by less than two degrees of freedom.
Other explicit constraints can originate from fixed singularities of the equation, requiring analyticity conditions (e.g.  \cite{Comellas:1997tf}), and it is hoped that the equation does not have too many such fixed singularity, which would require an over constraining of the solutions \cite{Benedetti:2012dx,Dietz:2012ic}. 

We will apply the following strategy to select solutions: i) we look for singularities of the equations, either fixed or movable, and study the behavior of the solution in a neighborhood of the singularity,  ii) we study the large field asymptotic solutions of the equation and count the degrees of freedom of each class of solutions,
iii) we numerically look for global solutions satisfying all the constraints.

The study of the large field asymptotic solutions is important also for other two reasons, namely, the derivation of the full effective action at the fixed point \cite{Benedetti:2012dx},
and the relation to the $f(R)$ theory, as we will explain in the concluding section.

We will present most of the analysis for the case $\om=0$, although occasionally we will refer to other values. 
As in the Landau gauge the $\om=0$ value is a critical value, analogous to the $\om=-1/2$ value for the Feynman gauge, we will treat separately the two gauges, starting with the Feynman gauge.
Most of our considerations apply to generic dimension $d>2$, although we will most often specialize to $d=4$. In Appendix \ref{App:2d} we will treat the special case $d=2$.

\subsection{Feynman gauge}
\label{Sec:anal:Feynm}

\subsubsection{Fixed singularities}
\label{Sec:anal:Feynm:fix-sing}

In order to look for fixed singularities, we search for poles of the denominator of the scale invariant flow equation $\partial_t \, \Vt_k = 0$ written in normal form, i.e.
\be \label{normal}
\Vt''(\phit)= \f{\cN(\Vt,\Vt',\phit)}{\cD(\Vt,\Vt',\phit)}\, ,
\ee
where $\cN$ and $\cD$ are polynomial functions obtained from  (\ref{RGFeyn}).
For $d>2$ the only zero we find is at $\phit = 0$, while for $d = 2$ the equation reduces to a first order equation with no fixed singularities. 
To test the consequences of such singularity in $d>2$ we impose analyticity, and study the equation in a Laurent expansion.

Locally, imposing analyticity means requiring the existence of a Taylor expansion of the solution, in other words we make the ansatz $\Vt(\phit) = \sum_{n\geq 0} v_n \phit^n$, and after plugging it into the equation we expand the latter in a Laurent series centered at the origin.
At leading order, the equation in the Feynman gauge reduces to 
\be
\label{a1}
0 = (2\, \om+1) \left(\frac{ 2}{2^d\, \pi^{d/2}\, d^2\, v_0\, \Gamma(d/2) + 4\, d}  - 1\right)\, ,
\ee
which vanishes either restricting to $\omega = - 1/2$ 
(the analogous case in Landau gauge will be $\omega = 0$, see \ref{Sec:anal:Landau:fix-sing}), 
or fixing the potential in the origin to
\be
v_0\equiv \Vt(0) = - \frac{2^{1-d}\, (2\, d - 1)}{\pi^{d / 2}\,d^2\, \Gamma(d/2)}\, .
\ee
As a consequence for $d>2$ and $\om\neq-1/2$ we have one constraint, thus reducing the number of degrees of freedom at the origin to one.
For technical reasons, when integrating the equation numerically, we need to start from an arbitrary small value of the field $\epsilon$. The boundary condition at $\epsilon$ can then be parametrized in terms of the derivative of the field in zero 
\be
\Vt(\epsilon) = \Vt(\epsilon; \Vt(0), \tau), \hspace{1cm} \Vt'(\epsilon) = \Vt'(\epsilon; \Vt(0), \tau)\, ,
\ee
being $\tau = \Vt'(0)$ the free parameter, and evaluated by means of a MacLaurin series
\be
\Vt(\epsilon) = -\frac{2^{1-d} (2\, d-1) \pi^{-d/2}}{d^2\, \Gamma(d/2)} + \tau\,  \eps + v_2(\tau) \, \eps^2 + 
\mathcal{O}(\eps^3)\, ,
\ee
where for example
\be
v_2(\tau) = \frac{2^{d}\, d\, \pi^{d/2}\, \Gamma(d/2)\,\left \{d^2\, ((d-1)\, d\, (2\, \omega + 1) - 4\,  \omega ) - 2\, d^2\, \tau^2 - 4\, (d - 2)\, (2\, d - 1) (2\, \omega +1) \tau \right \}}{8 \,(d-2) (2\, d-1)}\, ,
\nonumber
\ee  
and higher order coefficients are likewise obtained solving the equation order by order in $\eps$.

\subsubsection{Movable singularities}
\label{Sec:anal:Feynm:mov-sing}
The constrained differential equation admits now a one parameter family of local solutions parametrized by $\tau$. Still, because of the non linearity of the equation, we expect most of the solutions to end at a movable singularity, i.e. at a singularity whose location depends on the initial condition. We want to study the behavior of solutions in the neighborhood of such singularities, in order to confirm analytically the existence of such singularities and be able to recognize them in the numerical integrations, as well as to discuss possible interpretations in the terms of the $f(R)$ theory.
We will present in the next section the results of our search for a set of values of $\tau$ for which the singularity goes to infinity.  

Let $\phit_c$ be the value of the field at which the singularity occurs, and suppose that the singular behavior is such that there exists an $n_0\geq 0$ such that $\Vt^{(n)}(\phit_c)\sim\infty$ for every $n\geq n_0$.
In order to understand what values of $n_0$ can occur for our equation, it is convenient to recast the equation (\ref{RGFeyn}) in the following form
\be
\label{RGFeynrecast}
- d\, \Vt(\phit) + (d - 2)\, \Vt'(\phit)\, \phit  + \frac{1}{2^d\, d\, \pi^{d/2} \Gamma(d/2)}\, \frac{P_1(\Vt'', \Vt', \Vt, \phit)}{P_2(\Vt'', \Vt', \Vt, \phit)}   = 0\, ,
\ee
where the $P_i$ are two polynomials containing the same monomials but with different coefficients.
As the polynomials  $P_i$ have the same structure we deduce that for $\phit \to \phit_c$ their ratio will in general go to a constant for any value $n_0$.
Special situations can arise when  some cancellation occurs in $P_2$ which does not happen in $P_1$, and such cases will have to be discussed separately.
As a consequence, in the general case the linear part of the equation cannot diverge, otherwise it could not be balanced by the rational part, i.e. both the potential and its first derivative do not diverge at the singularity, restricting the possible value of $n_0$ to $n_0>1$.
At this stage, we can assume that in the neighborhood of $\phit_c$ the potential can be written as
\be
\label{critb}
\begin{split}
\Vt(\phit) = & (\phit - \phit_c)^\gamma \left \{A + A_1\, (\phit - \phit_c) + \mathcal{O}((\phit - \phit_c)^2) \right \} \\
& + u_0 + u_1 \,  (\phit-\phit_c)  + \mathcal{O}((\phit-\phit_c)^2) \, ,
\end{split}
\ee
and that $\g>1$ (so that $n_0>1$), and we can try to determine the value of $\gamma$ by means of the method of \textit{dominant balance}.
In order to do so we can start with the guess that the second derivative is divergent at $\phit_c$, that is $1<\g<2$.
In such case, by studying the balance of terms we arrive at the equation $\gamma-1 = - \gamma +2$, leading to $\gamma = 3/2$, in accordance with our guess.
Plugging \eqref{critb} with $\g=3/2$ into \eqref{RGFeynrecast},
we can iteratively work out all the coefficients in the expansion as functions of the parameter $u_0$ and of the singular field value $\phit_c$.
For example, in $d=4$ we find
\bea
u_1(u_0) &=& \frac{4 \,u_0\, \left(16\, \pi^2 (u_0 - \phit_c) + 1 \right) + \phit_c}{32\, \pi^2\, \phit_c\, (u_0 - \phit_c)} \, ,\\[1em]
A(u_0, u_1) &=& -\frac{\left (u_0^2\, (-(2\, \omega +1) ) + 2\, u_0\, (2\, \omega + 3)\, \phit_c + 2\,  u_1 \, \phit_c\,  \left(u_1 \, \phit_c - 2\,  u_0 \right) - (2\, \omega + 3)\,  \phit_c^2 \right)^{\frac{1}{2}}}{6\, \pi\, \phit_c\, (u_0 - \phit_c) \, \sqrt{2}}\, \nonumber ,
\eea
for the leading order terms. The subleading corrections can be computed iteratively, but their expression is very long, and not particularly enlightening.

Other singular behaviors are possible if $P_2$ has a zero. Such situations are more easily uncovered by studying the equation written in normal form, \eqref{normal}. 
Assuming that the first derivative of the potential is divergent (or more divergent than the potential itself) at $\phit\sim\phit_c$, we obtain the equation
\be \label{poleF}
\Vt''(\phit) \sim -2 \f{\Vt'(\phit)^2}{\phit_c-\Vt(\phit)} \, ,
\ee
leading to a simple pole solution $\Vt(\phit)\sim (\phit-\phit_c)^{-1}$, which is consistent with the assumption. Subleading corrections can be worked out, confirming the possibility that such type of singular behavior can appear in a solution of the fixed point equation.

\subsubsection{Behavior at large field values}
\label{Sec:anal:Feynm:large-fields}

We apply here the method of dominant balance to study the large field regime of the differential equation (\ref{RGFeyn}).
We have already seen in (\ref{RGFeynrecast}) that whatever is the leading term (for $\phit \to \infty$ in this case) the quantum part of the equation in general goes as a constant plus subleading corrections, hence we have two possibilities: either the potential diverges at infinity, and the classical part of the equation defines the leading order, or the potential goes to a constant, and there must be some balance between linear and nonlinear part.
In the first case, in the $\phit \to \infty$ limit the solution goes as 
\be
\label{asymp}
\Vt(\phit) \sim A\, \phit^\frac{d}{d-2} + \text{subleading terms} \, ,
\ee
where $A$ is a free parameter. Subleading terms can be calculated by solving iteratively the differential equation for an ansatz of the type
\be \label{infV}
\Vt(\phit) \sim A\, \phit^\frac{d}{d-2}\left(1 + \sum_{n>0} a_n(A)\, \phit^{-n}\right) \, .
\ee
For $d=4$, for example, the first few coefficients $a_n(A)$ are
\be
a_1(A) = 0\,, \hspace{1em} a_2(A) = -\frac{1}{16\, \pi^2}\, , \hspace{1em} a_3(A) = \frac{-2\, \omega -61}{1152\, \pi^2 A}\, , \hspace{1em} a_4(A) =\frac{-4\, \omega ^2-4\, \omega -337}{9216\, \pi ^2 A^2}\,  .
\ee
The coefficients are all inversely proportional to the bare parameter $A$, so that this expansion cannot be continued to $A = 0$, and that case must be treated separately.
The asymptotic solution so far constructed defines a one-parameter family of solutions parametrized by the variable $A$,
but as the equation is second order, we can ask if the asymptotic solutions have more degrees of freedom. 
In order to answer such question, we follow \cite{Morris:1994ki,Dietz:2012ic} and perturb the flow equation in the neighborhood of the solution we just found, i.e. we introduce a perturbation to the potential,
\be
\Vt(\phit) \to  \Vt(\phit)+ \epsilon\, \delta \Vt(\phit)\, ,
\ee
substitute it into (\ref{RGFeyn}), and expand to linear order in $\eps$.
Replacing $\Vt(\phit)$ with \eqref{infV}, and keeping only the leading terms in the coefficients of the linear operator acting on the perturbation, in $d=4$ we obtain the linear equation
\be
\label{marginal}
\frac{(-2\, \omega -1)\, \delta \Vt''(\phit)}{1152\, \pi^2\, A^2\, \phit} + 2\, \phit\, \delta \Vt'(\phit) - 4\, \delta \Vt(\phit) = 0\, ,
\ee
which allows a solution which goes asymptotically like
\be
\label{newdeg}
\delta \Vt(\phit) \sim B_1\, \phit^2 +  B_2 \, e^{\frac{768\, \pi^2\,A^2\,}{2\, \omega + 1}\, \phit^3}\, ,
\ee
where $B_1$ and $B_2$ are two integration constants. 
Note that eq. (\ref{marginal}) seems to reduce to a first order equation for $\omega = -1/2$, but as we will see for the Landau gauge for $\omega = 0$ (which is the analogue of the case $\omega = -1/2$ in the Feynman gauge) for that critical value of $\om$ we simply need to include the subleading correction of the coefficient of $\delta \Vt''(\phit)$.

Whereas the power-law solution in \eqref{newdeg} merely shifts $A$ in \eqref{infV}, the exponential solution would seem to be a new degree of freedom. However, for positive $\phit$  (and $\om>-1/2$, otherwise the role of positive and negative $\phit$ are interchanged) it grows faster than the solution it is perturbing, contradicting our asymptotic analysis, hence it must be discarded.
On the other hand, for negative $\phit$ it is an exponentially small perturbation, hence it is acceptable.
As the perturbation is smaller than any power at large $\phit$, while the leading solution  \eqref{infV} contains only powers, it is not difficult to see that the full equation decomposes in a hierarchy of equations, according to powers of the exponential correction, that is, the exponential acts like an $\eps$ parameter and we can iteratively solve the equation to obtain
\be
\Vt(\phit) \sim \sum_{m\geq 0} \left(B\,e^{Z(\phit)}\right)^m\, \Vt_{[m]} (\phit)\, ,
\ee
where $\Vt_{[0]}(\phit)$ is the leading solution \eqref{infV}, while for $\om=0$ we find
\be
Z(\phit) = 768 \,\pi^2 A^2 \phit ^3+4224\, \pi^2 A\, \phit^2+64 \left(24 \,A+769\, \pi^2\right) \phit\, ,
\ee
\be
\Vt_{[1]} (\phit) = \phit^{\frac{48 \left(329 \,A+5568\, \pi ^2\right)}{A}} \left(1- \frac{5712\, A^2+747937\, \pi ^2 A+8739072\, \pi ^4}{6\, \pi ^2 A^2} \phit^{-1} +     \mathcal{O}\left(\phit^{-2} \right) \right)\, ,
\ee
and so on, leaving $A$ and $B$ as free parameters.

The presence of a new degree of freedom at $\phit\sim-\infty$ creates an interesting situation, as we already know that we have an analyticity constraint at $\phit=0$, hence if we had just one-parameter families of solutions at both plus and minus infinity it would be unlikely to have a global solution.\footnote{Suppose that we start integrating at large positive $\phit$ with initial conditions dictated by  \eqref{infV}, and that reaching $\phit=0$ we compute $\Vt(0)$ and  $\t_+=\Vt'(0)$ as functions of $A$. Upon imposition of the analyticity condition we expect to find a discrete set of solutions for $A$. As $A$ was the only free parameter, $\t$ is now completely fixed by it. If at this point we repeat the same procedure but starting from large negative $\phit$, and if also in this case the asymptotic solutions form a one-parameter family, we will end up with a new fixed value $\t_-=\Vt'(0)$. It is very unlikely to find that $\t_+=\t_-$. On the contrary, if the asymptotic expansion at negative $\phit$ forms a two-parameter family, we could obtain in that case a continuum of values for $\t_-$, and
chances would be higher to find a global solution, as that would only require that $\t_+$ be in the range of $\t_-$.}

There remains to consider the special case $A=0$, which we now proceed to examine for $d=4$ and $\om=0$.
From the previous discussion of dominant balance we would expect in such case a solution that asymptotes to constant. Nevertheless, we should be careful as in that analysis we have excluded special cases leading to cancellations in the denominator of the quantum part of the equation.
By plugging into the equation an ansatz of the type
\be \label{infVspecial}
\Vt(\phit) \sim   A_1\, \phit +\sum_{n\geq 0} b_n\, \phit^{-n} \, ,
\ee
we find at leading order the equation $A_1=0$, in accordance with the previous analysis. However, a careful look at the higher orders of the expansion reveals the presence of poles at $A_1=1$ and $A_1=3/2$, meaning that for those values the general expansion is not valid, and a separate treatment is needed. In fact, we find that such special values of $A_1$ also lead to solutions that are solvable with an iterative algorithm.\footnote{For each of the special values of $A_1$ we find also poles in $b_0$, which however do not correspond to other solutions. Therefore we believe that we have exhausted the set of possible asymptotic solutions.}
In all three cases ($A_1=0$, $1$ and $3/2$) we find no free parameter in the expansion \eqref{infVspecial}, but by studying the linear perturbations we discover the presence of exponentially small corrections at negative $\phit$ for $A_1=0$, exponentially small corrections at both positive and negative $\phit$ for $A_1=1$, and a non-integer power correction at negative $\phit$ for $A_1=3/2$. It is quite easy to see that exponentially small corrections always carry one new degree of freedom, while the analysis in the case of the non-integer power is slightly more tedious and we have not pushed it further (also because in our numerical analysis we saw no evidence of the $A_1=3/2$ asymptotic behavior for the Feynman gauge).
Just as an example of the type of results, for $A_1=0$ we find that the coefficients in \eqref{infVspecial} read
\be \label{b-infVconst}
b_0 = \f{3}{128\,\pi^2}\, ,\;\;\;
b_1 = \frac{7}{6144\, \pi^4}\, ,\;\;\;
b_2 = \frac{985}{18874368\, \pi^6}\, ,\;\;\;
b_3 =\frac{4793}{1811939328\, \pi^8}\, ,
\ee
etc., and that the exponential perturbation at $\phit\sim-\infty$ leads to a solution of the form
\be
\Vt(\phit) \sim \sum_{m\geq 0} \left(B\,e^{192\, \pi^2 \phit}\right)^m\, \Vt_{[m]} (\phit)\, ,
\ee
where $\Vt_{[0]} (\phit)$ is the perturbed solution with coefficients \eqref{b-infVconst}, and
\be
\Vt_{[1]} (\phit) =  \phit^8\, \left(1-\frac{233}{128\, \pi ^2} \phit^{-1} + \mathcal{O}\left(\phit^{-2} \right) \right) \, ,
\ee
\be
\Vt_{[2]} (\phit) =  \phit^{17}\,  \left(6144\,\pi^4-\frac{463}{128 \,\pi ^2} \phit^{-1} + \mathcal{O}\left(\phit^{-2} \right) \right) \, ,
\ee
and so on, leaving $B$ as the only free parameter.

In conclusion, we found four isolated sets of solutions at $\phit\to \pm\infty$. As we will explain in the concluding section, from the point of view of the $f(R)$ theory the most interesting solutions are those in the first class, i.e. \eqref{asymp}, for which we have found the presence of two degrees of freedom at  $\phit\to-\infty$ and one at  $\phit\to+\infty$ (or the opposite for $\om<-1/2$).

\subsection{Landau gauge}
\label{Sec:anal:Landau}
\subsubsection{Fixed singularities}
\label{Sec:anal:Landau:fix-sing}
We repeat here the analysis of the analyticity of the differential equation for the Landau gauge, starting with the study of the fixed singularity in $\phit = 0$.
Following \ref{Sec:anal:Feynm:fix-sing}, we recast the differential equation in its normal form \eqref{normal} and then we expand it in a Laurent series employing a Taylor expansion for the potential.
In this gauge we find that at leading order the equation reduces to
\be
\label{a1landau}
0 =  - 4\, \omega \, \frac{2^d\, \pi^{d/2}\, d\, \Vt(0) \Gamma(d/2+1)+ d - 1}{d\, (2^d\, d\, \pi^{d/2} \,\Vt(0)\, \Gamma(d/2) + 2)} \, ,
\ee
which vanishes constraining the potential at the origin as
\be
v_0\equiv \Vt(0) = - \frac{2^{-d}\, (d-1)\, \pi^{-d/2}}{d\, \Gamma(d/2+1)}\,,
\ee
or restricting to $\omega = 0$, which is the case we are interested in. 
Comparing (\ref{a1landau}) with (\ref{a1}) we note once more that the case $\omega=0$ in the Landau gauge is analogous to the case $\omega= -1/2$ in the Feynman, so that the analytic properties of the equation in the two gauges are the same for those two particular values. 

For $\omega = 0$ we have now an equation free of singularities. As a consequence, since the equation is unconstrained, we have (for $d>2$) two degrees of freedom at the origin, $\Vt(0)$ and $\Vt'(0)$, and at least one at $\phit \pm \infty$, so that it seems more likely to find global solutions.
On the technical side, the absence of a singularity at $\phit=0$ also means that in this case it is possible to integrate numerically from the origin without employing a MacLaurin expansion.

\subsubsection{Movable singularities}
\label{Sec:anal:Landau:mov-sing}
As in the Feynman gauge we expect the non linearity of the equation to involve the presence of movable singularities.
Since the polynomials $P_i$ in equation (\ref{RGFeynrecast}) contain the same monomials in both gauges, 
the analysis carried out in \ref{Sec:anal:Feynm:mov-sing} with the method of the dominant balance still holds and we find in general the singular behavior \eqref{critb} with $\g=3/2$. However, because of the gauge dependence of the off-shell effective action, we end up with different coefficients for both the analytic and divergent part. For example, for $d=4$ and generic $\omega$ we obtain
\bea
\label{Landausingcoeff}
u_1(u_0) &=& \frac{1}{64}\, \left(\frac{128\, u_0}{\phit_c}+\frac{1}{\pi^2}\, \left ( \frac{5}{u_0 - \phit_c}+\frac{3}{2\, u_0 - 3\, \phit_c}+\frac{4}{\phit_c} \right)  \right) \, ,\\[1em]
A(u_0, u_1) &=& -\frac{(-3\, \phit_c^2 \left(6\, \omega - 4\, u_1^2 + 9 \right) + 12\, u_0\, \phit_c\, \left(2\, \omega - 2 \, u_1 + 3 \right) - 8\, \omega\,  u_0^2)^{\f12}}{6\, \pi \,\phit_c\, (3\, \phit_c - 2 u_0)\, \sqrt{2}} \, ,
\eea
etc.
Also similar to the Feynman gauge is the presence of simple pole singularities, with \eqref{poleF} replaced by
\be \label{poleL}
\Vt''(\phit) \sim -4\, \f{\Vt'(\phit)^2}{3\,\phit_c-2\,\Vt(\phit)} \, .
\ee

\subsubsection{Behavior at large field values}
\label{Sec:anal:Landau:large-fields}

Since the method of the dominant balance leads to similar conclusions for both gauge choices, we expect also for the Landau gauge to find generically an asymptotic solutions of the form
\be
\label{landauasymp}
\Vt(\phit) \sim A\, \phit^\frac{d}{d-2}\left(1 + \sum_{n>0} a_n(A)\, \phit^{-n}\right) \, .
\ee
We can iteratively solve the differential equation for this ansatz, obtaining in $d=4$
\be
\label{landaucorr}
a_1(A) = 0\, , \hspace{1em} a_2(A) = -\frac{1}{32\, \pi^2}\, , \hspace{1em}a_3(A) = \frac{-2 \omega - 39}{1152\, \pi^2 A}\, , \hspace{1em} a_4(A) = \frac{-4\, \omega^2-207}{9216\, \pi^2\, A^2}\, ,
\ee
and so on. As for the other gauge, we see that the coefficients are inversely proportional to $A$, so that also in this gauge we have to treat separately that case.
Before studying those other solutions we focus on the number of free parameters of (\ref{landauasymp}), by introducing a perturbation $\delta \Vt$. We then linearize the equation for the perturbation and study the leading terms, obtaining the equation
\be
\label{pertLandau1}
-\frac{\omega  \, \delta \Vt''(\phit)}{576\, \pi^2\, A^2\, \phit} + 2\, \phit \, \delta \Vt'(\phit) - 4\,\delta \Vt(\phit) = 0\, .
\ee
For $\om\neq 0$ the analysis is similar to the one we presented for the Feynman gauge.
For $\om=0$ we need to include the next order term in the coefficient of  $\delta\Vt''$, and thus consider the equation
\be
\label{pertLandau2}
\frac{\delta\Vt''(\phit)}{128\, \pi^2\, A^3 \phit^2} + 2 \,\phit \,\delta\Vt'(\phit) - 4 \,\delta\Vt(\phit) = 0\, ,
\ee
which admits solutions with the asymptotic behavior
\be
\label{perLandauexp}
\delta\Vt(\phit) \sim  \text{B}_1\,  \phit^2 + B_2\,  e^{- 64\, A^3\, \pi^2\, \phit^4}\, .
\ee
The novelty here is that the leading power in the exponent is fourth rather than third order (a consequence of the different power of $\phit$ in the coefficient of $\delta\Vt''$ in \eqref{pertLandau2} with respect to \eqref{pertLandau1}), so that the solution does not discriminate positive from negative $\phit$, but rather leads to constraints on $A$.
For $A<0$, the solution (\ref{perLandauexp}) contains an exponential degree of freedom which grows faster then the perturbed function in both positive and negative field regimes, so that we must discard it. Interestingly such sector is the unphysical one, since negative $A$ defines the asymptotic behavior of an unbounded potential. 
On the other hand, for $A>0$ the perturbation is exponentially small both at positive and negative $\phit$, hence it is always acceptable, and we can work out the subleading corrections as done before for the Feynman case. 
The higher power in the exponent means that we have to solve more iteration steps before getting to the power-law corrections, but as we do not gain any qualitative insight from such analysis, we do not report further on that, the main message being that now we have two degrees of freedom at both plus and minus infinity.

Regarding the case $A=0$, making the ansatz \eqref{infVspecial} we find again ($d=4$ and $\om=0$) the same three special values $A_1=0$, $1$ and $3/2$, as in the Feynman gauge.
The main difference appears in the case $A_1=3/2$, for which the expansion \eqref{infVspecial} now contains one degree of freedom, i.e. $b_1$ is a free parameter in terms of which all the other $b_n$ are expressed:
\be \label{b-infV32}
b_0 =- \f{3}{64\,\pi^2}\, ,\;\;\;
b_2 = -\frac{b_1}{8\, \pi^2}\, ,\;\;\;
b_3 =\frac{b_1 (11-1024\, \pi^4 b_1)}{1024\, \pi^4 }\, ,\;\;\;
\text{etc.}
\ee
By perturbing around such solution we find that in order to discover new solutions we have to include at least the next-to-leading order coefficients for large $\phit$ in the linear equation, yielding 
\be
\left(\frac{64\, \pi^4 b_1 -1}{2\, \pi^2 b_1} \phit ^3-\frac{\phit^4}{b_1}\right)
   \delta\Vt''(\phit)+\left(-\frac{512\,\pi^4 b_1 -3}{4\, \pi ^2 b_1} \phit ^2-\frac{2\, \phit^3}{b_1}\right)
   \delta\Vt'(\phit)+\left(64\, \pi^2 \phit +\frac{4}{3}\right) \delta\Vt(\phit) = 0 \, ,
\ee
whose asymptotic solutions are a superposition of a solution that simply perturbs \eqref{b-infV32}, and a series of logarithmic corrections,
\be
\delta\Vt(\phit) \simeq c_1 \log \phit \left( \phit^{-1} - \f{\phit^{-2}}{8\pi^2}   + \mathcal{O}\left(\phit^{-3} \right)  \right)\, ,
\ee
that carries a second degree of freedom, namely the free parameter $c_1$.

\section{Numerical results}
\label{Sec:fps}
In order to find global solutions we integrate out from $\phit=0$ and search for a set of initial conditions $\tau$ such that the movable singularity goes to infinity in both the positive and negative field region. 
We present here our analysis for both gauges for $\omega = 0$ and $d=4$, starting with the Feynman gauge.

\subsection{Feynman Gauge}
\label{Sec:fps:Feynman}
We start a numerical integration at the origin (actually at $\phit=\pm\eps$ as explained in Sec.~\ref{Sec:anal:Feynm:fix-sing}), and similarly to what done in \cite{Morris:1994ki}, we plot the location at which we hit a singularity, as a function of the free parameter $\t = \Vt'(0)$. When we see a spike in such a plot, we interpret it as a hint of a possible global solution. 
Since spikes can occur as artifacts due to the scale of the plot, ending instead at a finite value, the next step is to show that such spike can be made arbitrarily long by increasing the numerical precision and by refining the mesh. In addition, in our case we have to produce such type of plots at both positive and negative $\phit$, looking for spikes that occur at the same value of $\t$ in both ranges.

\begin{center}
\begin{figure}[ht]
\begin{center}
\includegraphics[width=8cm]{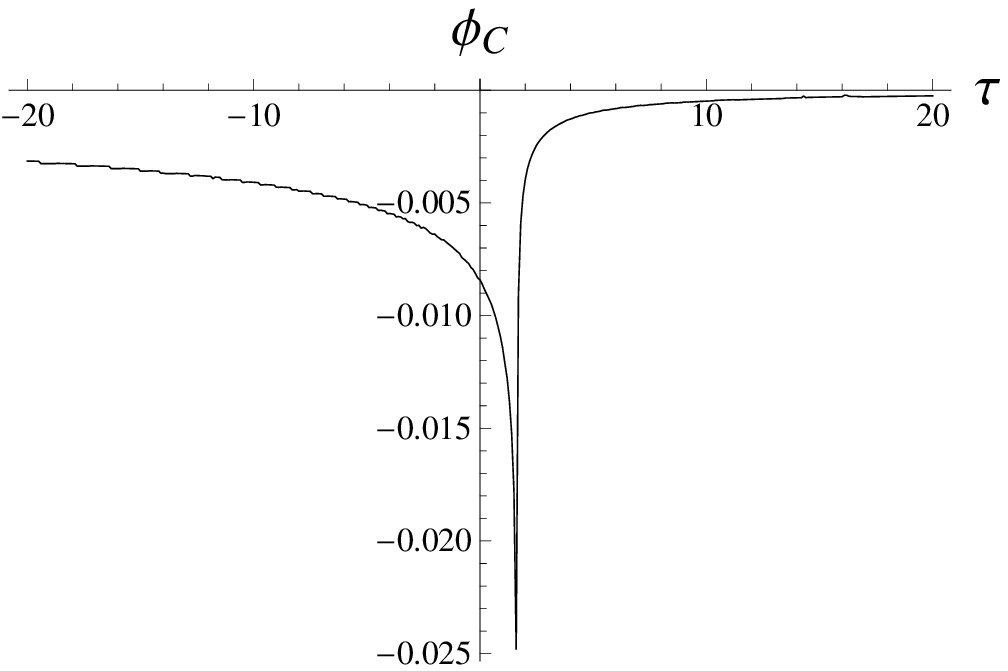}
\includegraphics[width=8cm]{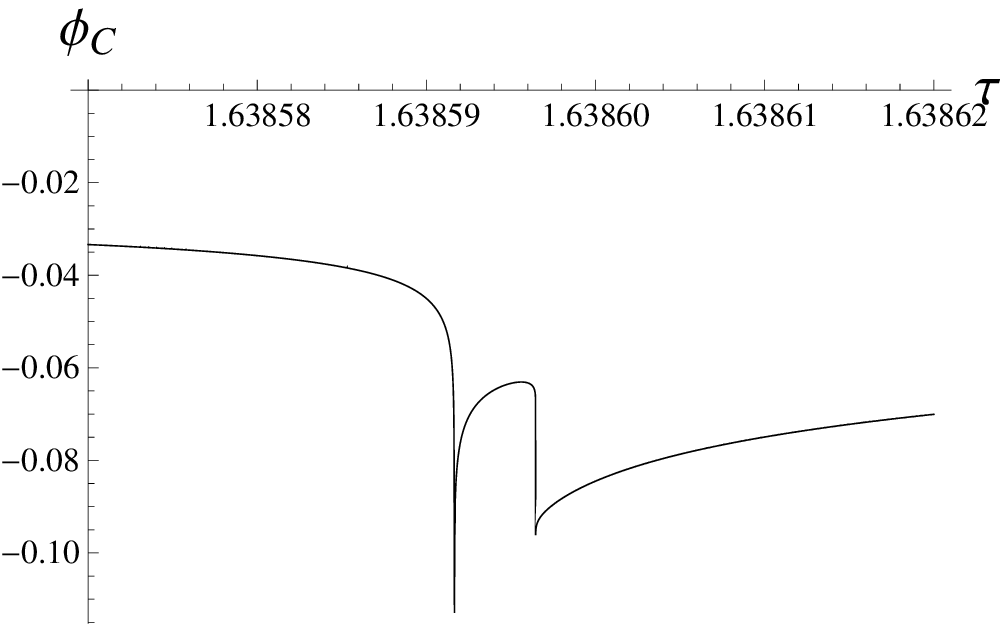}
\caption{
The critical field value $\phi_c$ in the negative domain, as a function of the initial condition $\tau = \Vt'(0)$, for $d = 4$ and $\omega = 0$ in the Feynman gauge.
In the right panel is a blow up of the spike, showing two spikes discussed in the text. A third spike at  $\t\sim 1.638534$ is not shown here.}
\label{Fig:d4om0neg}
\end{center}
\end{figure}
\end{center}

At negative $\phit$ the plot of the singularities looks like in Fig.~\ref{Fig:d4om0neg}.
We apparently find a spike in the negative region for an initial condition $\tau \sim 1.638$, which however, when zooming in, reveals a richer fine structure, actually three peaks being present (only two of which are shown in the right panel of Fig.~\ref{Fig:d4om0neg}).

\newpage

\begin{figure}
\centering
\begin{tabular}{cc}
\includegraphics[width=7.5cm]{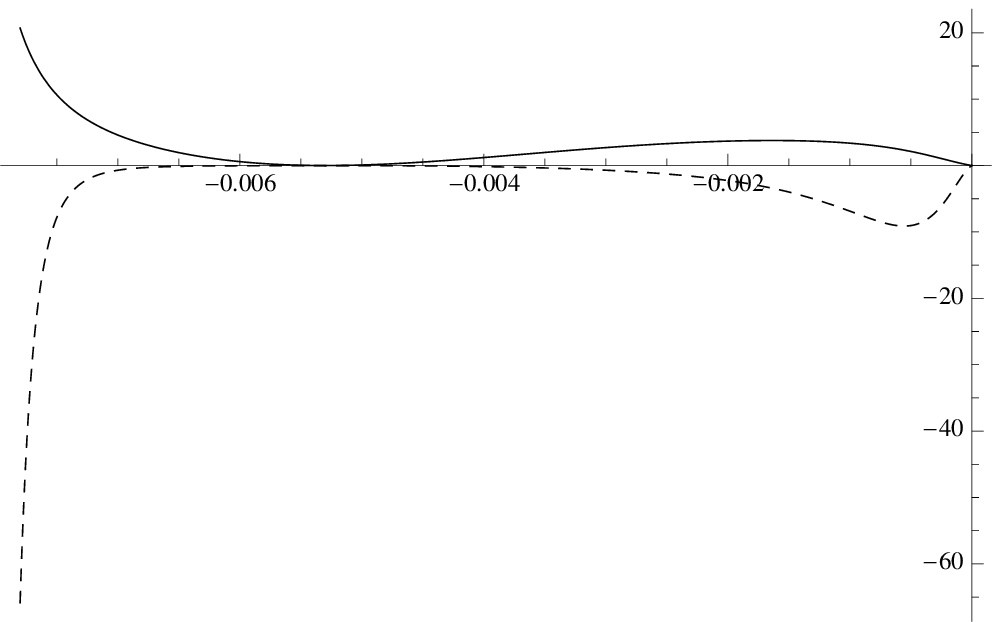} & \includegraphics[width=7.5cm]{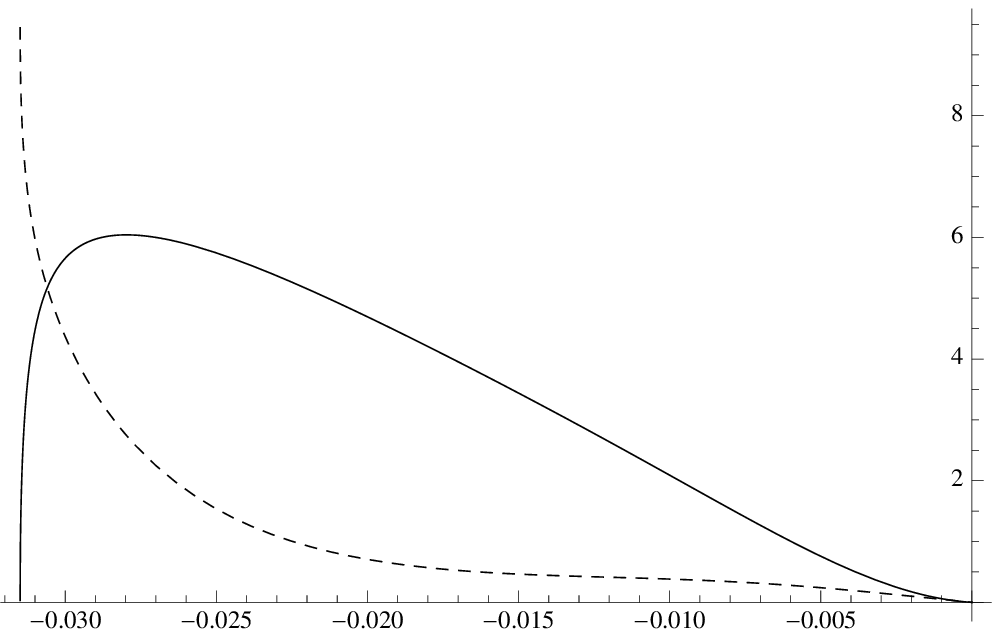} \\
\includegraphics[width=7.5cm]{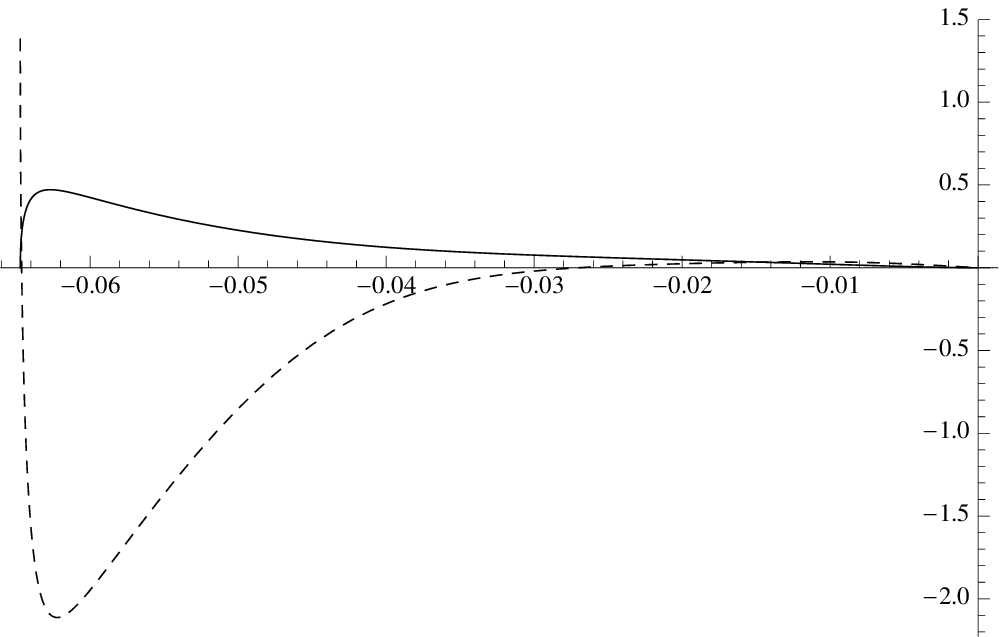} & \includegraphics[width=7.5cm]{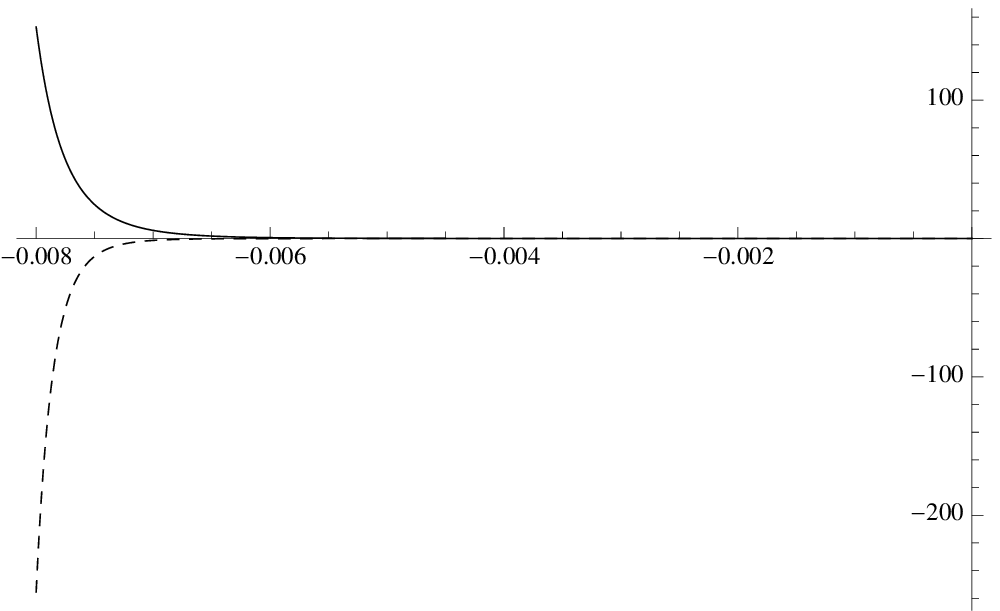} 
\end{tabular}
\caption{A table of plots of $\cN$ (dashed) and $\cD$ (solid) as functions of $\phit$, at four values of $\t$ (from top to bottom, left to right, $\t=-5$, $\t=1.63855$, $\t=1.6385965$ and $\t=1.7$) corresponding to the four different regimes we observed when integrating at negative $\phit$. Plots are not to scale, typically $\cD$ is several orders of magnitude smaller than $\cN$. 
}
\label{Fig:negNDplotsF}
\end{figure}

\begin{center}
\begin{figure}[ht]
\begin{center}
\includegraphics[width=8cm]{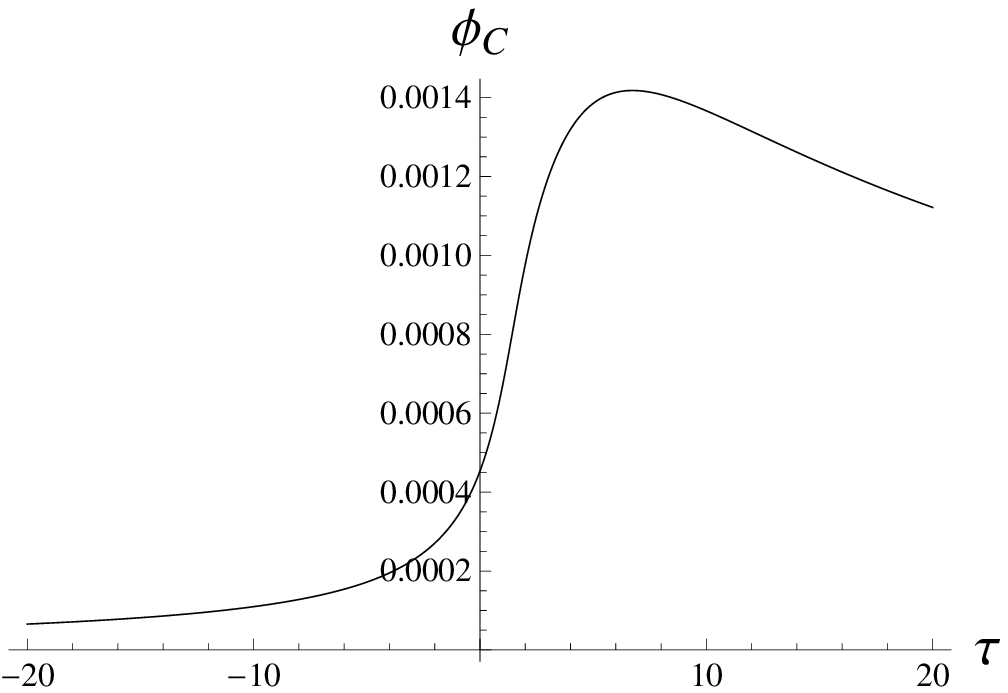}
\caption{
The critical field $\phi_c$ in the positive domain, as a function of the initial condition $\tau = \Vt'(0)$, for $d = 4$ and $\omega = 0$ in the Feynman gauge.}
\label{Fig:d4om0pos}
\end{center}
\end{figure}
\end{center}

\begin{center}
\begin{figure}[ht]
\includegraphics[width=8cm]{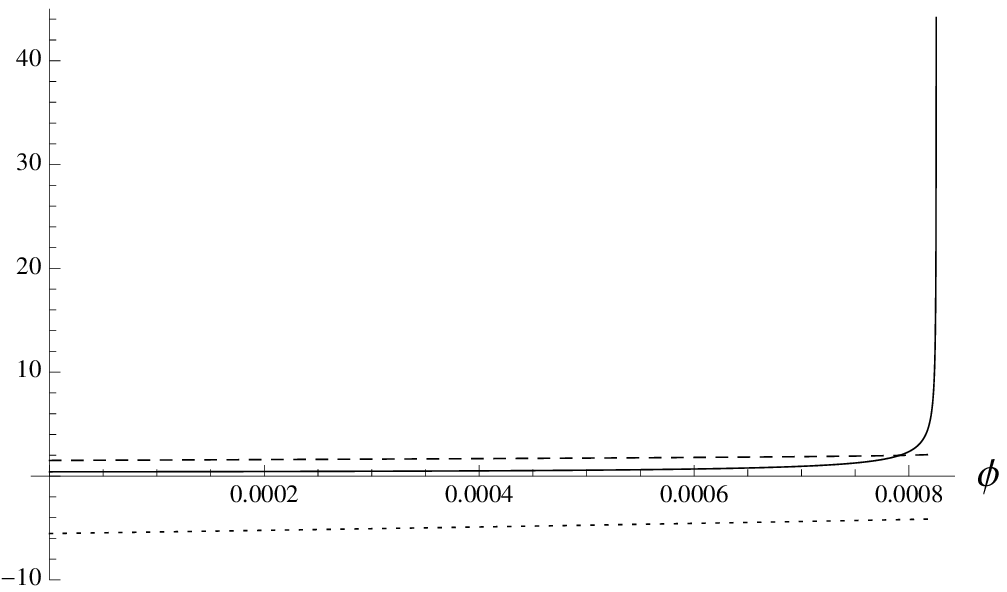}
\includegraphics[width=8cm]{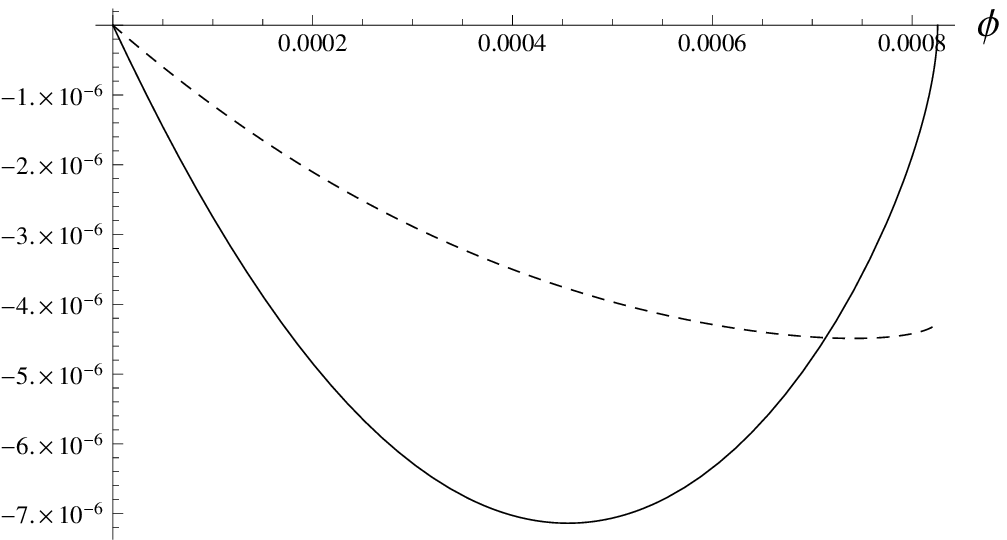}
\caption{
Typical plots of solutions hitting a singularity in the positive domain. We show here the case $d=4$, $\omega=0$ for Feynman gauge with $\t=1.5$. 
The left panel shows the potential (rescaled by a factor $10^3$) together with its first and second derivative (rescaled by a factor $10^{-3}$), respectively in dotted, dashed and continuous lines.
The right panel shows the behavior of $\cN$ (dashed) and  $10^3\times \cD$.
}
\label{Fig:d4om0singsolF}
\end{figure}
\end{center}

Such triple peak can be understood in terms of transition between different types of singular behavior. The most clear explanation is obtained in terms of the numerator and denominator of the normal equation, $\cN$ and $\cD$ in \eqref{normal}, which we plot in Fig.~\ref{Fig:negNDplotsF} for four representative cases.
We find that for $\t\lesssim 1.638534$ and $\t\gtrsim 1.638597$ both $\cN$ and $\cD$ diverge, together with their ratio, at some $\phit_c$ thus signaling the pole type of singularity found in  \eqref{poleF}. In the range between those two value we find that $\cD$ vanishes  at some $\phit_c$, reaching zero with an infinite slope; at the same $\cN$ reaches a finite value, and we deduce that we are hitting a singularity of the type \eqref{critb} with $\g=3/2$. 
The transitions between $\g=-1$ and $\g=3/2$ coincide with two of the peaks observed in the fine structure of Fig.~\ref{Fig:d4om0neg}. We interpret the remaining spike at $\t\sim 1.638591$ as signaling a transition (as $\t$ increases) from a regime in which $\cN$ is always positive, to one in which it changes sign twice before hitting hitting $\phit_c$. 
As seen in the zoomed plot in Fig.~\ref{Fig:d4om0neg}, spikes can be pushed farther away from the origin, however, high precision is needed and we have not tried to reach much beyond $\phit_c\sim-0.1$.
In fact, it turns out that a more detailed investigation of the spikes is not worth, as the remaining part of the plot, for positive $\phit$, turns out to be quite disappointing. 
Integrating in the positive field region for any initial condition, including the neighborhood of  $\tau \sim 1.638$, we encounter a singularity, as can be seen in Fig.~\ref{Fig:d4om0pos}, so that we would have not in any case a global solution. Only one type of singular behavior is found in the positive domain,
a typical example of which is shown in Fig.~\ref{Fig:d4om0singsolF}, and
from which we recognize a behavior  consistent with \eqref{critb} and $\g=3/2$.

We did not find other spikes in both negative and positive region for other values of $\tau$ (outside the plot range in Fig.~\ref{Fig:d4om0pos}), so that in the end we conclude that there are no global solutions in $d=4$ and $\omega=0$ in the Feynman gauge.

\subsection{Landau Gauge}
\label{Sec:fps:Landau}
The search of global solutions is more complicated in the Landau gauge since we have two degrees of freedom at the origin. In order to search for fixed points we adopted the following strategy: i) we integrate numerically from the origin 
(since there is no fixed singularity we can directly impose initial conditions at $\phit=0$) for a fixed value of $\Vt(0)$ varying the initial condition $\tau = \Vt'(0)$, ii) we repeat the integration for a discrete set of positive and negative values of $\Vt(0)$. As for the Feynman gauge we restrict our research to $\omega = 0$ and $d=4$.
\begin{center}
\begin{figure}[ht]
\begin{center}
\includegraphics[width=9cm]{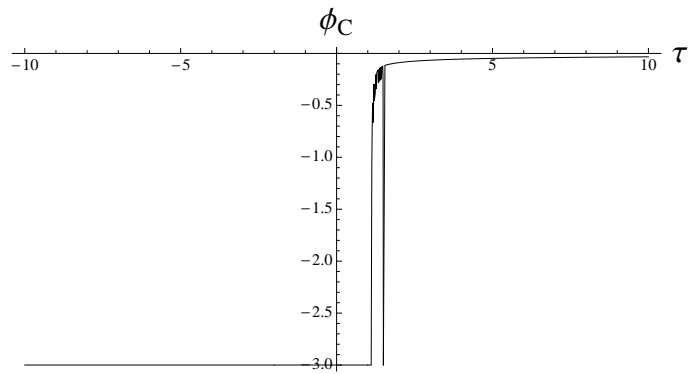}
\caption{
The critical field $\phi_c$ as a function of the initial condition $\tau = \Vt'(0)$ for $\Vt(0) = 0.1$, $d = 4$ and $\omega = 0$ in the Landau gauge.}
\label{Fig:d4om0Lpos}
\end{center}
\end{figure}
\end{center}
We start with $\Vt(0)>0$, for which we illustrate a representative outcome at negative $\phit$ in Fig.~\ref{Fig:d4om0Lpos}. In this case we find a spike at  $\tau = 1.5$ and a continuum set of analytic solutions occurring for $\tau < \tau_c$, where $\tau_c$ is a critical value which depends on the initial condition $\Vt(0)$, i.e. $\tau_c \equiv \tau_c(\Vt(0))$.
The peak at $\tau = 1.5$  actually corresponds to an exact solution of the differential equation in normal form, which for generic $d>2$ is given by the simple linear function 
\be
\label{Landaulinear}
\Vt(\phi) = A + \frac{2\,(d-1)}{d} \phit \, ,
\ee
being $A= \Vt(0)$ a free parameter. However, we should be careful about such solution, as in the original equation it corresponds to a zero of both numerator and denominator of the $h$-$\varphi$ trace, leading to an undetermined expression. The reason for the zero in the denominator is easily found by looking back at the second variation \eqref{HessLand}, and taking $\om=0$ and a linear function for $V(\phi)$: the $\varphi$-$\varphi$ component immediately vanishes, while the $h$-$\varphi$ component does so once we implement the rule \eqref{cutrule} in combination with \eqref{optimized} and we choose the linear function as in \eqref{Landaulinear} (the $h$-$h$ component vanishes only for $A=0$). As a consequence, the $h$-$\varphi$ matrix is not invertible in such case. We also cannot use a limiting procedure to attribute to \eqref{Landaulinear} the status of solution of the original equation, as writing $\Vt(\phi) = A + \frac{2\,(d-1)}{d} \phit +\eps\, v(\phit)$ and expanding in $\eps$ we find that the zeroth order term in $\eps$ does not vanish, leading instead to a nonlinear differential equation for $v(\phit)$ (implying also that \eqref{Landaulinear} does not admit linear perturbations).
We are thus led to deem \eqref{Landaulinear} unacceptable.

Regarding the continuum set at negative $\phit$, we find it for an initial conditions $\tau$ smaller then a critical value $\tau_c$ which, as we already mentioned, depends on the value of the initial condition $\Vt(0)$. Varying $\Vt(0)$ we observed the value of $\tau_c$ to oscillate between a minimum value $\tau_{min} \sim 0.96$ and a maximum $\tau_{max} \sim 1.12$. 
By increasing the numerical precision we were able to prolong at will the entire group of solutions and we found all of them to behave asymptotically as $A\, \phit^2$, being $A$ a function of the initial conditions. 
\begin{center}
\begin{figure}[ht]
\begin{center}
\includegraphics[width=10cm]{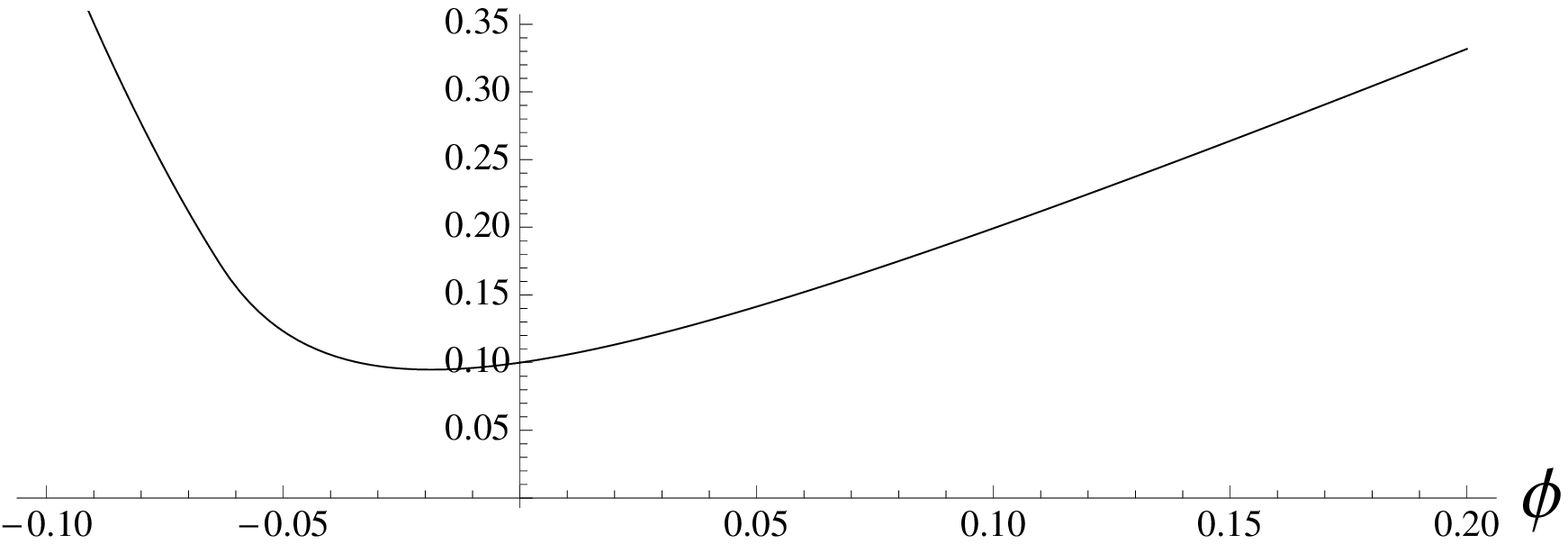}\\[-1em]
\includegraphics[width=10cm]{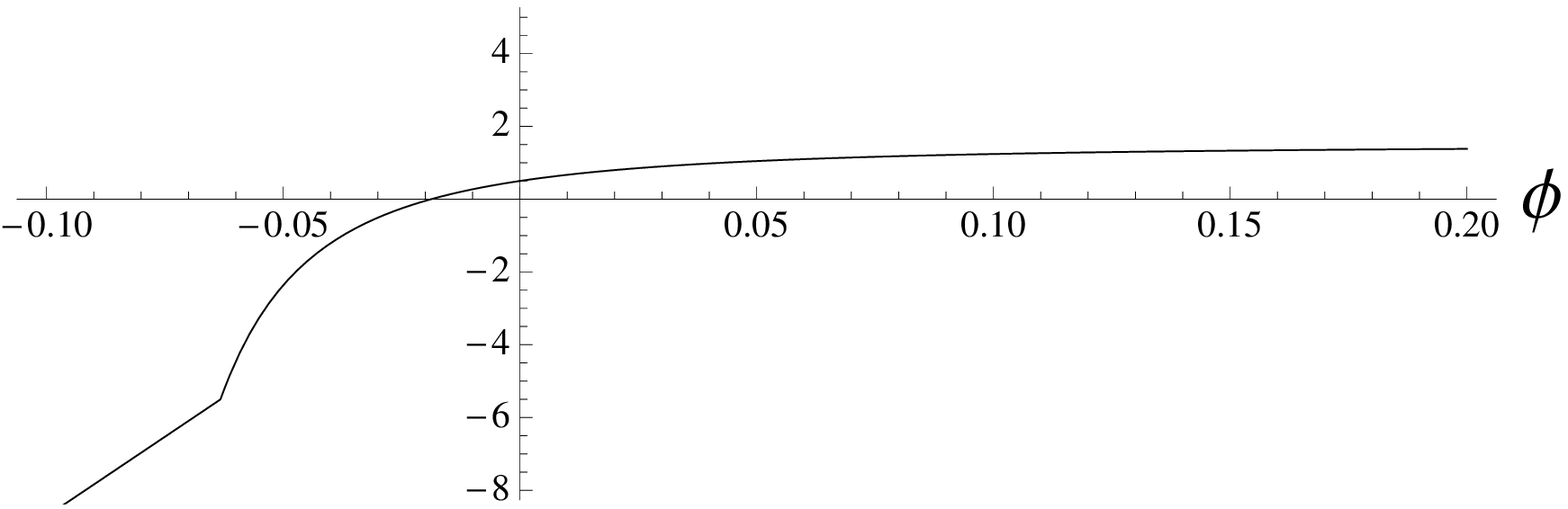}\\[-1em]
\includegraphics[width=10cm]{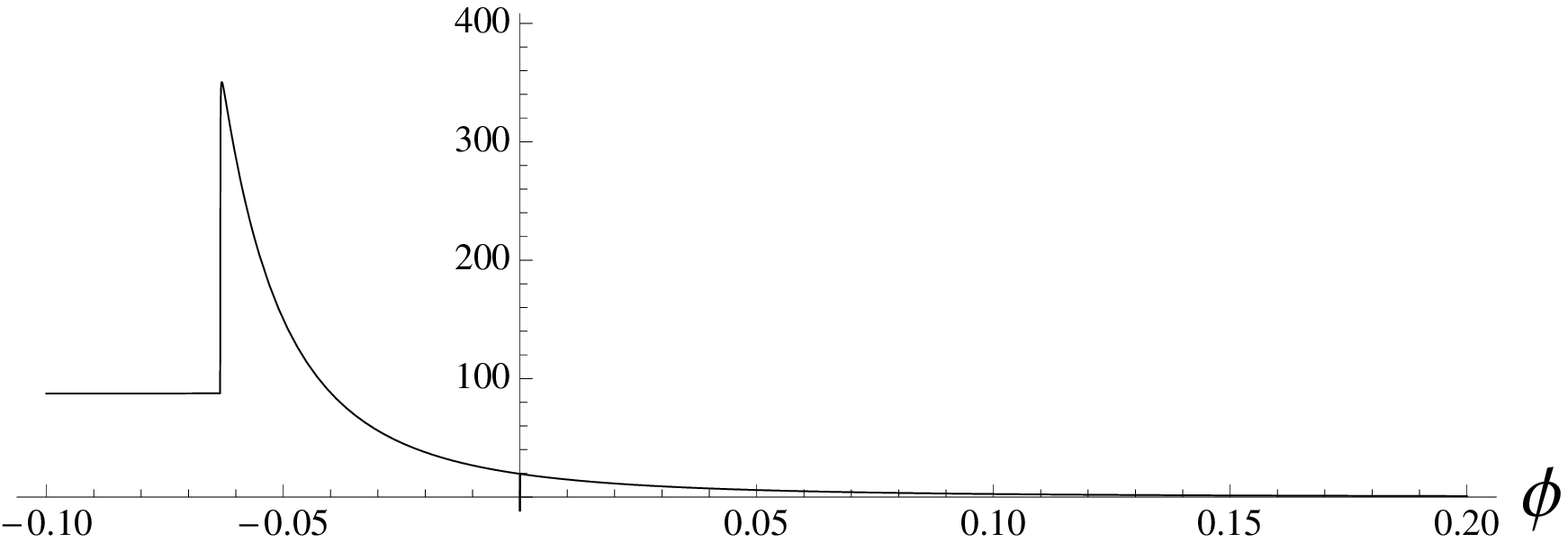}
\caption{
Plot of a typical global solution in the Landau gauge ($\tau = 0.5$, $\Vt(0) = 0.1$, $d = 4$ and $\omega = 0$). 
In the upper, central and bottom panel are plotted respectively the potential, its first and its second derivative.
}
\label{Fig:d4om0Lpos2}
\end{center}
\end{figure}
\end{center}

\begin{center}
\begin{figure}[ht]
\begin{center}
\includegraphics[width=8cm]{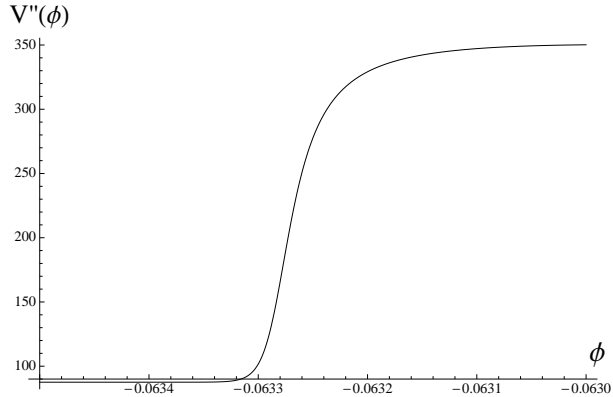}
\caption{
Plot of the second derivative of the potential in the range of the exponential transition to the asymptotic solution $\Vt(\phit) \sim A\, \phit^2$ in the Landau gauge ($\tau = 0.5$, $\Vt(0) = 0.1$, $d = 4$ and $\omega = 0$). 
}
\label{Fig:d4gobba}
\end{center}
\end{figure}
\end{center}

A typical solution is illustrated in Fig.~\ref{Fig:d4om0Lpos2}. 
The seemingly sharp edge in the second derivative is actually an optical artifact: working at high precision, and zooming around the edge one finds that the curve is smooth, as depicted in Fig.~\ref{Fig:d4gobba}.
We can understand the presence of such a short-scale transition as the rapid vanishing at large $\phit$ of the exponential part of the solutions we discussed in Sec.~\ref{Sec:anal:Landau:large-fields} (it can be deduced from Fig.~\ref{Fig:d4om0Lpos2} that $A>0$, hence the exponential corrections are possible). The flat tail in the second derivative is also well understood in terms of the power-law asymptotic solution (\ref{landauasymp}-\ref{landaucorr}), according to which (in $d=4$) the second derivative of the potential approaches a constant plus $O(1/\phi^3)$ corrections (with coefficient of the $1/\phi^3$ correction being in this case of the order of $10^{-4}$, whereas the constant term is of order $10^2$).

As it can be seen in Fig.~\ref{Fig:d4om0Lpos} all the numerical integrations performed using initial conditions with $\tau > \tau_c$ lead (with the exception of $\tau = 3/2$) to a singularity, which we found to be characterized by the exponent $\gamma = 3/2$. An acurate analysis reveals a transition in the way the solutions behave before reaching the movable singularity (i.e. the large field regime of the solution), from $\Vt(\phit) \sim A\, \phi^2$ at $\tau \sim \tau_c$, to $\Vt(\phit) \sim \f32 \, \phit$ at $\tau \sim 3/2$. 
Such transition, together with the spurious solution \eqref{Landaulinear}, makes the equation particularly stiff around $\tau = 3/2$, as it can be seen from the noise in Fig.~\ref{Fig:d4om0Lpos}. 
However, because of the presence of a singularity we did not put much effort on a more precise numerical integration of this group of solutions.

Integrating towards positive $\phit$ we discover an interesting situation: for $\Vt(0)>0$ no solutions meet any singularity.
We were able to push the integration to arbitrarily large $\phit>0$ without encountering singularities for all values $\tau$, and we found solutions with $\tau < 3/2$ to behave asymptotically like $\Vt(\phit) \sim \f32 \phit$, and solution with $\tau > 3/2$ to go as $\Vt(\phit) \sim A\, \phit^2$.
Combining our findings for positive and negative $\phit$ we conclude that the solutions with $\Vt(0)>0$ and  $\tau < \tau_c$ form a continuous set of global solutions.

At $\phit=0$ and  $\Vt(0) = 0$ the equation is singular. Imposing an analyticity condition at the origin we find that $\t=(1\pm\sqrt{19})/4$. We did not study these special solutions in detail.

\begin{center}
\begin{figure}[ht]
\begin{center}
\includegraphics[width=10cm]{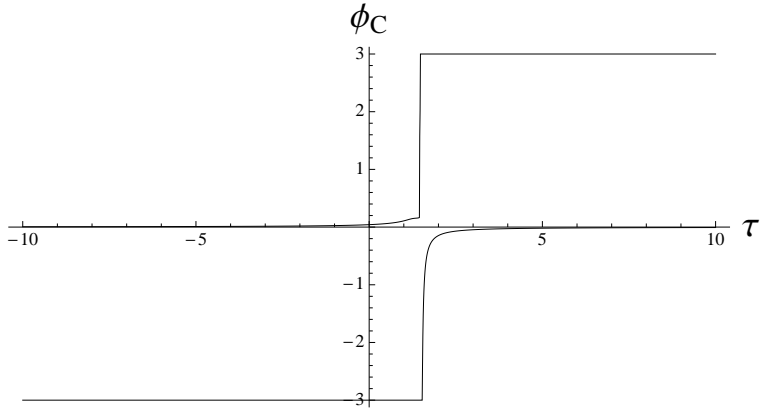}
\caption{
The critical field $\phi_c$ as a function of the initial condition $\tau = \Vt'(0)$ for $\Vt(0)= - 0.1$, $d = 4$ and $\omega = 0$ in the Landau gauge.}
\label{Fig:d4om0Lneg}
\end{center}
\end{figure}
\end{center}

For $\Vt(0) < 0$ the typical situation is depicted in Fig.~\ref{Fig:d4om0Lneg}.  All the singular solutions we found, for both positive and negative field values, diverge with exponent $\gamma = 3/2$. 
We found in the positive field region a continuum of solutions which do not end on a movable singularity for $\tau > 3/2$, while at negative $\phit$ we met no singularity for $\tau < 3/2$, in both cases with an asymptotic behavior  $\Vt(\phit) \sim \f32 \phit$. The two sets have no overlap, hence there are no global solutions in this case.

In conclusion, in the Landau gauge in $d=4$ and $\om=0$, we found a two parameter family of global solutions for $\Vt(0)>0$ and $\t<\t_c(\Vt(0))$. Such result could have been expected to some extent, as in the Landau gauge we have no fixed singularity at the origin, and we have at least two classes of asymptotic behavior with two degrees of freedom each at both positive and negative $\phit$.
The global solutions we found behave asymptotically as  $\Vt(\phit) \sim A\, \phit^2$ for $\phit\to-\infty$, and as $\Vt(\phit) \sim \f32 \, \phit$ for $\phit\to+\infty$.
The latter is an indication of an unusual character of such solutions, as that type of asymptotic behavior is the result of a balance between the classical and quantum parts of the RG equation, to be contrasted to the usual situation, where for $k\to 0$ (i.e. the large field regime) only the classical part survives.

\section{Conclusions}
\label{Sec:concl}

In this article we have presented a study of the Brans-Dicke theory \eqref{BD-action} for an arbitrary potential $V(\phi)$ in the framework of the functional renormalization group. We have derived a differential equation in the local potential approximation for a generic parameter $\omega$ and dimension $d$, subsequently focusing our analysis on the case $\omega = 0$ and $d = 4$, because of its classical equivalence with the $f(R)$ theory.
The main motivation for this work came from the asymptotic safety scenario of quantum gravity, which in the literature has been investigated mostly in the pure metric formulation, by means of truncations of the exact renormalization group equations. An important test for such approximation methods would be to show that at least some subclasses of truncations correspond to a series expansion of a functional approximation that explores an infinite dimensional subspace of the full theory space, a classical successful example of that being the local potential approximation in scalar field theory. However, for the case of gravity, such direction is progressing slowly because of the notorious difficulties related to working on curved backgrounds.
The idea we have explored in this paper was to exploit the classical equivalence between Brans-Dicke theory and $f(R)$ gravity in order to be able to study a functional approximation of gravity on a flat background.
Besides such motivation, a study of alternative formulations of gravity or modified theories is interesting in its own, and studies like the one we presented here can help address the question of quantum equivalence of different formulations.

In order to achieve our goals we have evaluated the renormalization group equation for a generic potential $\Vt(\phit)$ in two different gauge choices, namely a Feynman and a Landau gauge,  allowing us to discern possible gauge artifacts.
As a result of our study, we found a number of important differences between the two gauges.
In particular, we found no global solutions of the fixed point equation in the Feynman gauge, whereas we found a two-parameter family of global solutions in the Landau gauge.
While some gauge dependence was expected (due to the approximations employed and to the fact of working off-shell, see for example \cite{Falkenberg:1996bq,Benedetti:2011ct}),  we would have expected that at least qualitative features like the number of fixed points, and of associated relevant directions would be gauge independent (in principle together with any observable quantity, but in practice this property is expected to hold only approximately due to the approximations used). 
Being the results in our two gauges so different even at a qualitative level, we  are led to infer some inconsistency of the model under consideration in the present approximation.
Motivated by the relation to $f(R)$ gravity, we did not analyze the case $\omega \neq 0$ in detail, but we can identify the freezing of the Brans-Dicke parameter to $\om=0$ as the culprit of the inconsistent scenario we uncovered.
We expect the strong gauge dependence to be lifted once the Brans-Dicke parameter is promoted to a running coupling  $\om_k$, in the sense that in any gauge there will be some critical value $\om_c$ where something special happens (e.g. a discrete or continuous set of fixed points appears), the value of $\om_c$ being gauge dependent, but not so the overall picture.\footnote{Our hypothesis is partially supported by the $d=2$ case, in which the flow equations turn out to be $\om$-independent, and give similar results in the two gauges. See Appendix \ref{App:2d}.} For example, we already know that in the Feynman gauge the value $\om=-1/2$ gives very similar results to the Landau gauge at $\om=0$, and it would be interesting to test whether such critical values correspond to  fixed points of $\om_k$ for the two gauges, reached either in the UV or in the IR.
In view of our results and of the possible solution we just outlined, we can draw an important conclusion: due to its renormalization group flow, the Brans-Dicke theory at the quantum level needs a running coupling $\om_k\neq 0$ in order to be consistent. Since for $\om\neq 0$ the equivalence with the $f(R)$ theory is broken, we are led to suggest that Brans-Dicke theory and $f(R)$ gravity are inequivalent at the quantum level. Needless to say, this should not be intended as a proof of inequivalence, but rather as a logical interpretation of the results we found.

We should point out another aspect which also hints to a non-equivalence of  Brans-Dicke theory and $f(R)$ gravity at the quantum level.
As we explained, the condition for a solution of the FRGE to be a valid fixed point is that it should be a global solution.
While it is quite clear from our analysis that, at least within the present approximation, no nontrivial fixed point can be found for the Brans-Dicke theory at $\om=0$ in the Feynman gauge, we should be careful in translating such statement back into $f(R)$ gravity. Due to the nonlinearity of the Legendre transform it could happen that a problematic singularity in one theory would turn into a harmless one in the other, or vice versa. We should indeed remember that the following relations hold (here in dimensionless variables):
\be \label{R-phi}
\Rt=\Vt'(\phit)\, ,\;\;\; \ft'(\Rt)=-\phit\, .
\ee
As a consequence, if a singular point $|\phit_c|<\infty$ is such that the first derivative of the potential is divergent, then in the $f(R)$ theory it simply means that $\phit_c$ is mapped to $\Rt_c=\pm \infty$, depending on the sign of $\Vt'(\phit_c)$. Although that would correspond to a strange situation in which $\ft'(\Rt)$ does not diverge at infinity (usually the asymptotic behavior is a power law dictated by the tree level part of the equation \cite{Benedetti:2012dx,Dietz:2012ic,Benedetti:2013jk}, implying that at infinity  $\ft'(\Rt)$ diverges for any $d>2$), that would not be something we can discard as unacceptable. This is precisely what happens in reverse for the Landau gauge: we found global solutions for $\Vt(\phit)$, but their first derivative is such that asymptotically $\Vt'(\phit)\sim3/2$ for $\phit\to+\infty$, and thus their transform would lead to an $f(R)$ theory valid only up to $\Rt_c=3/2$.
On the other hand, if the potential is such that only its derivatives of order greater or equal to two are divergent, then the singular point is mapped to $|\Rt_c|<\infty$, and thus also the transform of the potential is not a global function. The latter is precisely the case for the Feynman gauge, for which we saw that the singularities at positive $\phit$ are characterized by an exponent $\g=3/2$, that is, they have a finite first derivative at the singular point.

Regardless of its connection to the $f(R)$ approximation, the study of Brans-Dicke theory is interesting in its own, and, being a nonrenormalizable theory, it is natural to wonder whether an asymptotic safety scenario applies to it. 
From such point of view, we should emphasize that what we have presented here is the result of the leading order in an approximation which should be systematically improved. 
The local potential approximation we employed can be considered, in fact, as a ``double LPA'' since we neglected both the renormalization of the coupling $Z$ of the operator $\phi\, R$ (having set from the start $Z=1$) and of the parameter $\omega$. Both could be promoted to functions $Z(\phi)$ and $\om(\phi)$, thus leading to a next-to-leading order approximation which could uncover an anomalous scaling of $\phi$ and the existence of nontrivial fixed points.

\subsection*{Acknowledgements}
%
We thank Roberto Percacci for useful remarks and corrections on the first version of the paper.

\newpage
\appendix
\section{The two-dimensional case}
\label{App:2d}

In two dimensions the fixed point equations in both gauges reduce to  $\om$-independent first order equations. The analysis is thus quite different in this case, it is actually much easier, and we can proceed mostly by analytical means.

Explicitly, the equations in $d=2$ reduce to\footnote{The field is dimensionless in two dimensions, hence we omit the tilde.}
\be
\label{2dFeyn}
\frac{\Vt(\phi)+ (\phi-2\, \Vt(\phi)) \Vt'(\phi)}{2\, \pi\,  (\phi-\Vt(\phi)) \left(1-\Vt'(\phi)\right)}-2\, \Vt(\phi) = 0 \, ,
\ee
for the Feynman gauge, and to
\be \label{2dLandau}
\frac{\Vt'(\phi)}{2\, \pi\,  \left(1-\Vt'(\phi)\right)} -2 \Vt(\phi)  = 0\, ,
\ee
for the Landau gauge.
Both equations can be easily integrated, leading to algebraic equations implicitly defining the solution $\Vt(\phi)$.
As equation \eqref{2dFeyn} is slightly more complicated to study than equation  \eqref{2dLandau}, but at the end it leads to very similar results, we will present the explicit analysis only for the Landau gauge.
The fact that the two gauges lead to similar results in this $\om$-independent case supports our hypothesis that in higher dimensions the strong gauge dependence we found is an artifact of the restriction to $\om=0$.

Equation \eqref{2dLandau} can be integrated to give the algebraic relation
\be
\Vt - y_0 +\f{1}{4\pi} \log\left(\Vt/y_0\right) = \phi - \phi_0 \, ,
\ee
whose solution is by definition expressed in terms of the Lambert function $\mathcal{W}$,
\be \label{2dLambert}
\Vt(\phi) = \frac{\mathcal{W}\left(4\, \pi\,  e^{C + 4\pi  \phi}\right)}{4\, \pi\, } \, .
\ee
The constant of integration $C=4\,\pi\, (v_0-\phi_0) + \log y_0$ parametrizes a one-parameter family of global solutions, which hence are all acceptable fixed points.
The asymptotic behavior of the Lambert function is such that $\Vt(\phi)\sim \phi$ for $\phi\to+\infty$, and $\Vt(\phi)\sim e^{4\pi \phi+C}$ for  $\phi\to-\infty$.

We can study the linear perturbations around such fixed points, by writing as usual
\be
\Vt_k(\phi) = \Vt(\phi) + \eps\, v(\phi)\, e^{-\l t} \, ,
\ee
with $\Vt(\phi)$ given by \eqref{2dLambert}. Expanding to first order in $\eps$, we find the eigenvalue equation
\be
(2-\l) \,v(\phi) = \frac{\left(1+ \mathcal{W}\left(4\, \pi\,  e^{C+4 \pi  \phi}\right)\right)^2}{2\, \pi } \, v'(\phi) \, ,
\ee
whose solutions are
\be
v(\phi) =A \left(\frac{\mathcal{W}\left(4\, \pi\,  e^{C+4 \pi  \phi}\right)}{\mathcal{W}\left(4\, \pi\,  e^{C+4 \pi \phi}\right)+1}\right)^{\frac{2-\lambda }{2}} 
  = A \left( \Vt'(\phi) \right)^{\frac{2-\lambda }{2}}   \, , 
\ee
$A$ being an arbitrary normalization constant, which we can fix to one.
Given the exponential fall-off at $\phi\sim-\infty$ of the fixed point solution $\Vt(\phi)$, we see that we must impose the constraint $\l\leq 2$ in order to avoid exponentially growing perturbations.
Indeed the asymptotic behavior of the eigenperturbations is $v(\phi)\sim 1-\f{2-\l}{2} (4\pi\phi)^{-1}$ for $\phi\to+\infty$, and $v(\phi)\sim (4\pi)^{\frac{2-\lambda }{2} }\,e^{\frac{2-\lambda }{2}(4\pi \phi+C)}$ for  $\phi\to-\infty$.
Apart from the upper bound on $\l$, we do not have other restrictions, hence the perturbations form a continuous spectrum. 
However, for $\l< 2$ all the perturbations are redundant, corresponding to a field redefinition $\phi\to \phi + (  \Vt'(\phi))^{-\l/2}$.
We are left with only one essential perturbation, the constant one, $v(\phi) =A$.

One special solution of the fixed point equation is $\Vt(\phi)=0$, whose linear perturbations satisfy
\be
(2-\l) \,v(\phi) = \frac{1}{2 \,\pi } \, v'(\phi) \, ,
\ee
with solutions $v(\phi) = A \, e^{(2-\l) 2\pi \phi}$. In order to avoid exponentially growing solutions in this case we have to restrict to $\l=2$, that is, the only allowed perturbation is again a constant potential, which is a relevant perturbation, and which actually is an exact solution of the full flow equation.

We conclude noting that all the solutions in $d=2$ do not admit an $f(R)$ interpretation, as \eqref{R-phi} together with the asymptotic behavior of the fixed point solutions imply that $\Rt\in (0,1)$. The departure from $f(R)$ is of course most evident in the $\Vt(\phi)=0$ case, where the equation of motion obtained by varying $\phi$ is simply $R=0$.

\providecommand{\href}[2]{#2}\begingroup\raggedright\endgroup

\end{document}